\documentclass[journal]{IEEEtran}
\usepackage{setspace}
\usepackage{graphicx,bm}
\usepackage{epstopdf} 
\usepackage{amsmath}
\usepackage{color}
\usepackage{amsfonts}
\allowdisplaybreaks[4]
\usepackage{bm,bbm}
\usepackage{cite}
\usepackage{verbatim} 

\newcounter{bla}
\newtheorem{assumption}{Assumption}
\newtheorem{theorem}{Theorem}

\begin{document}

\title{Testing the Structure of a Gaussian Graphical Model with Reduced Transmissions in a Distributed Setting}

\IEEEoverridecommandlockouts
\author{Yicheng Chen, Rick S. Blum,
\IEEEmembership{Fellow,~IEEE}, Brian M. Sadler, \IEEEmembership{Fellow,~IEEE}, and Jiangfan Zhang, \IEEEmembership{Member,~IEEE}
\thanks{The work is supported by the U. S. Army
Research Laboratory and the U. S. Army Research Office under grant number
W911NF-17-1-0331.}
\thanks{
Yicheng Chen and R. S. Blum are with Lehigh University, Bethlehem, PA 18015 USA
(email: yic917@lehigh.edu, rblum@eecs.lehigh.edu).}
\thanks{
Brian M. Sadler is with the Army Research Laboratory, Adelphi, MD 20878 USA (email: brian.m.sadler6.civ@mail.mil).}
\thanks{
Jiangfan Zhang is with the Department of Electrical and Computer Engineering, Missouri University  of Science and Technology, Rolla, MO 65409 USA
(email: jiangfanzhang@mst.edu). }}
 \IEEEoverridecommandlockouts

\maketitle

\begin{abstract}
Testing a covariance matrix following a Gaussian graphical model (GGM) is considered in this paper based on observations made at a set of distributed sensors grouped into clusters. Ordered transmissions are proposed to achieve the same Bayes risk as the optimum centralized energy unconstrained approach but with fewer transmissions and a completely distributed approach. In this approach, we represent the Bayes optimum test statistic as a sum of local test statistics which can be calculated by only utilizing the observations available at one cluster. We select one sensor to be the cluster head (CH) to collect and summarize the observed data in each cluster and intercluster communications are assumed to be inexpensive. The CHs with more informative observations transmit their data to the fusion center (FC) first. By halting
before all transmissions have taken place, transmissions can be saved without performance loss. It is shown that this ordering approach can guarantee a lower bound on the average number of transmissions saved for any given GGM and the lower bound can approach approximately half the number of clusters when the minimum eigenvalue of the covariance matrix under the alternative hypothesis in each cluster becomes sufficiently large.
\end{abstract}

\begin{IEEEkeywords}
Covariance matrix testing, distributed detection, energy efficiency, Gaussian graphical models, ordered transmission.
\end{IEEEkeywords}

\IEEEpeerreviewmaketitle

\section{Introduction}

Great attention has been devoted to distributed detection in sensor networks for both military and civilian
applications, such as security, tactical surveillance, defense operations, disaster prediction and health care monitoring \cite{akyildiz2002wireless}. However, since spatially distributed sensor nodes have limited battery capacity, energy efficiency is an important topic \cite{appadwedula2005energy}\cite{6047550}\cite{4545248}\cite{zhang2017ordering}.
Recently, for the shift-in-mean detection problem, \cite{4545248} proposed an approach where transmissions can be ordered and halted before all transmissions have taken place based on the assumption that the observations are independent and identically distributed. The work in \cite{zhang2017ordering} has generalized the ordered transmission approaches to mean-shift detection with statistically dependent observations. However, covariance matrix testing problems employing ordered transmissions have not been addressed yet. This paper differs from the previous work on ordered transmissions for the shift-in-mean problem \cite{4545248}\cite{zhang2017ordering}, and focuses on problems testing the covariance matrix \cite{wei2012distributed}\cite{aittomaki2011resource} by employing a Gaussian Graphical Model (GGM) formulation.

A GGM characterizes dependencies using a graph with a node for each variable and edges labeled with weights that describe the correlation between the connected nodes. The observations of each node are assumed to follow a Gaussian distribution and the dependency between the observations is represented by the edge. Since GGMs can be used to describe high dimensional parametric models with a small number of samples via message passing algorithms, they have received extensive study in speech recognition \cite{bilmes2005graphical}, Internet backbone networks \cite{wiesel2009decomposable}, genetic networks \cite{wille2004sparse}, image processing \cite{willsky2002multiresolution}, machine learning \cite{meng2013distributed}\cite{mohan2014node}, sensor networks \cite{cetin2006distributed} \cite{guestrin2004distributed} and electrical power systems \cite{wei2012distributed}\cite{he2011dependency}\cite{electric6687941}. In this paper, we employ the GGM to organize the sensors into clusters based on the largest fully connected subgraphs they belong to, and this provides the possibility of applying ordered transmissions.


We consider testing the graphical structure of a decomposable GGM with reduced transmissions in a distributed scenario.  We first group all sensors into several clusters. These clusters correspond to the largest fully connected subgraphs (called cliques). Then we show we can decompose the global optimum test statistic into a sum of local test statistics, each of  which only depends on the observations in the corresponding cluster. Each cluster selects one sensor as the cluster head (CH) which we assume will have more computation and communication capacity \cite{abbasi2007survey}. The other sensors in the cluster transmit their observations to the CH and then the CH calculates the local test statistic and transmits it to the fusion center (FC). Typically the sensors in a cluster (clique) are located close to each other\footnote{The correlation structure implied by a clique will usually imply close physical proximity in the applications we have surveyed, see \cite{he2011dependency} for example.}, so local communications within the cluster do not use as much energy as the communications from the CH to the FC. We employ ordered transmissions over the CHs to reduce the more costly communications from the CHs to the FC. Specifically, after collecting and summarizing the observations from other sensors in the cluster, the CH sets a timer to decide when to transmit its local test statistic. All CHs are assumed to be synchronized and the timer in each cluster is inversely proportional to the magnitude of the local test statistic so that the CH with the most informative observations will transmit first, followed by the next most informative CH, and so on. This provides a method to apply ordered transmissions to solve covariance matrix testing problems with dependent observations. Developing an ordering algorithm for covariance matrix testing problems is one contribution of this work. The second major contribution is that a lower bound on the average number of transmissions saved is derived in this paper which is a valid lower bound for all cases we consider. The last contribution is that we have shown that when the minimum eigenvalue of the covariance matrix under the alternative hypothesis in each cluster becomes sufficiently large, nearly half of the transmissions can be omitted which provides a limiting behavior of the general lower bound on the number of transmissions saved.

\subsection{Notation and Organization}

Throughout this paper, bold lower case letters are used to denote column vectors, and bold upper case letters denote matrices. For any vector ${\bf x} \in {\mathbbm{R}}^N$ and any set ${\cal S} \subseteq \{1,2,...,N\}$, the vector ${\bf x}_{\cal S}$ is defined as a subvector of $\bf x$ which consists of the elements of $\bf x$ corresponding to the indices contained in $\cal S$. For any matrix ${\bf A}$, let ${( {\bf{A}} )_{i,j}}$ denote the element in the $i$-th row and $j$-th column. ${\bf{A}} \succ 0$ and ${\bf{A}} \succeq 0$ imply that the matrix $\bf A$ is positive definite and positive semidefinite, respectively. The eigenvalues of $\bf A$ are denoted as $eig\{{\bf A}\}$.  Given any two sets of indices ${\cal U}$ and ${\cal V}$ where ${\cal U} \subseteq {\cal V}$, let $\bf A$ denote a matrix whose coordinates correspond to the indices in ${\cal U}$,
then $[ ({\bf A}_{\cal U})^{-1}]^{\cal V}$ denotes the matrix obtained from the appropriate zero-filling needed to obtain a dimension $|{\cal V}|$-by-$|{\cal V}|$ matrix as per
\begin{equation}\label{defineVoperation}
{\bigg[ ({\bf A}_{\cal U})^{-1}\bigg]^{\cal V}_{i,j}} = \left\{ {\begin{array}{*{20}{l}}
	{{{\left( {\bf{A}}^{-1} \right)}_{i,j}}},\; &{\text{ if } i \in {\cal U},\; j \in {\cal U}}\\
	0,\; & \text{ otherwise. }
	\end{array}} \right.
\end{equation}
For example, if we let ${\bf{A}}^{-1}=[1,1;1,1]$, ${\cal U}=\{2,3\}$ and ${\cal V}=\{1,2,3\}$, then we obtain $[ ({\bf A}_{\cal U})^{-1}]^{\cal V}=[0, 0, 0;0, 1, 1;0, 1, 1]$.

The rest of paper is organized as follows. The basics of decomposable GGMs, the covariance testing problem and the distributed test statistic are presented in Section \ref{Covchange123}. Ordering for covariance matrix testing problems and a lower bound on the average number of transmissions saved via ordering are provided in Section \ref{reducedtrans}. Performance analysis and numerical examples are presented in Section \ref{numericalresu}. Finally, conclusions are drawn in Section \ref{conclusion}.

\section{Covariance Matrix Testing in Decomposable Gaussian Graphical Models}\label{Covchange123}

In this section, we introduce the properties of decomposable GGMs and formulate the multivariate Gaussian covariance matrix testing problem in a distributed scenario.

\subsection{Decomposable Gaussian Graphical Models}
Consider an undirected graph ${\cal G} = \left( {{\cal V},{\cal E}} \right)$ where ${\cal V} = \{1,2,...,N\}$ is the set of indices and ${\cal E} =\{(i_1,j_1),(i_2,j_2),...,(i_{|{\cal E}|},j_{|{\cal E}|})\}$ denotes the set of undirected edges of the graph.
Let a random vector ${\bf{x}} \buildrel \Delta \over = {\left[ {{x_1},{x_2},...,{x_N}} \right]^T}$ follow a multivariate Gaussian distribution which also satisfies the Markov property with respect to ${\cal G}$ which implies that if $(i,j) \notin {\cal E}$, then
\begin{equation}
{\left( {{{\bf{\Sigma }}^{ - 1}}} \right)_{i,j}} = 0
\end{equation}
where ${\bf{\Sigma }}$ is the covariance matrix of $\bf x$, and ${{{\bf{\Sigma }}^{ - 1}}}$ is referred to as the concentration matrix (also known as the information matrix). Note that the topology of the graph $\cal G$ is described by the non-zero elements of the concentration matrix ${{{\bf{\Sigma }}^{ - 1}}}$.

An undirected graph is decomposable if it can be successively decomposed into its cliques \cite{lauritzen1996graphical}. Throughout the paper, we concentrate on decomposable undirected graph models.  Let $K$ denote the number of cliques in the decomposable undirected graph ${\cal G}$. The perfect sequence of cliques of the graph ${\cal G}$ is denoted by $\{{\cal C}_1, {\cal C}_2,...,{\cal C}_K\}$. We denote the corresponding histories $\{{\cal H}_k\}_{k=1,2,...,K}$ and separators $\{{\cal S}_k\}_{k=2,3,...,K}$ as
\begin{equation} \label{Define_histories}
{{\cal H}_k} = {{\cal C}_1} \cup {{\cal C}_2} \cup  \cdots  \cup {{\cal C}_k}, \; \forall k=1,2,...,K,
\end{equation}
and
\begin{equation} \label{Define_separators}
{{\cal S}_k} = {{\cal H}_{k - 1}} \cap  {{\cal C}_k}, \; \forall k=2,...,K.
\end{equation}
Note that for all $k>1$, there is a $j<k$ such that ${\cal S}_k \subseteq {\cal C}_j$ for any decomposable undirected graph $\cal G$.

A mapping $q: \{ {2,3,...,K} \} \to \{ {1,2,....,K} \}$ is defined to specify an association between each separator set and some unique cliques such that \cite{zhang2017ordering}
\begin{equation} \label{Define_q_k}
q\left( k \right) \buildrel \Delta \over = \min \left\{ {j \left| \; {{{\cal S}_k} \subseteq {{\cal C}_j}} \right.} \right\}, \; \forall k=2,3,...,K.
\end{equation}
Thus, the $k$-th separator ${\cal S}_k$ is associated with the $q(k)$-th clique ${{\cal C}_{q(k)}}$ according to
\begin{equation} \label{S_k_C_k}
{{\cal S}_k} \subseteq {{\cal C}_{q(k)}}.
\end{equation}
Note that the $k$-th separator ${\cal S}_k$ is not only contained in the $q(k)$-th clique ${\cal C}_{q(k)}$, but also contained in the $k$-th clique ${{\cal C}_k}$ as defined in (\ref{Define_separators}), that is,
\begin{equation} \label{S_k_C_k1}
{{\cal S}_k} \subseteq {{\cal C}_k}.
\end{equation}
It is worth mentioning that for any $k>1$, $q(k)$ must exist and
 \begin{align}\label{q_k_smaller_k}
 q(k)<k
\end{align}
for any decomposable undirected graph $\cal G$. Let ${\cal Q}_j$ denote the set of indices of the separators which are associated with the $j$-th clique via the mapping $q$ in (\ref{Define_q_k}), that is,
\begin{equation} \label{Define_Qj}
{{\cal Q}_j} \buildrel \Delta \over = \left\{ {k\left| \; {q\left( k \right) = j} \right.} \right\}.
\end{equation}
From (\ref{q_k_smaller_k}), we know that the minimal element in ${{\cal Q}_j}$ satisfies
\begin{equation}
\min {{\cal Q}_j} > j,
\end{equation}
which implies that
\begin{equation} \label{Q_j_subset}
{{\cal Q}_j}  \subseteq \left\{ {j + 1,j + 2,...,K} \right\}, \; \forall j=1,2,...,K-1,
\end{equation}
and
\begin{equation}\label{emptydefin}
{{\cal Q}_K}  = \emptyset.
\end{equation}

\begin{figure}[!t]
\centering
\includegraphics[width=2.5in]{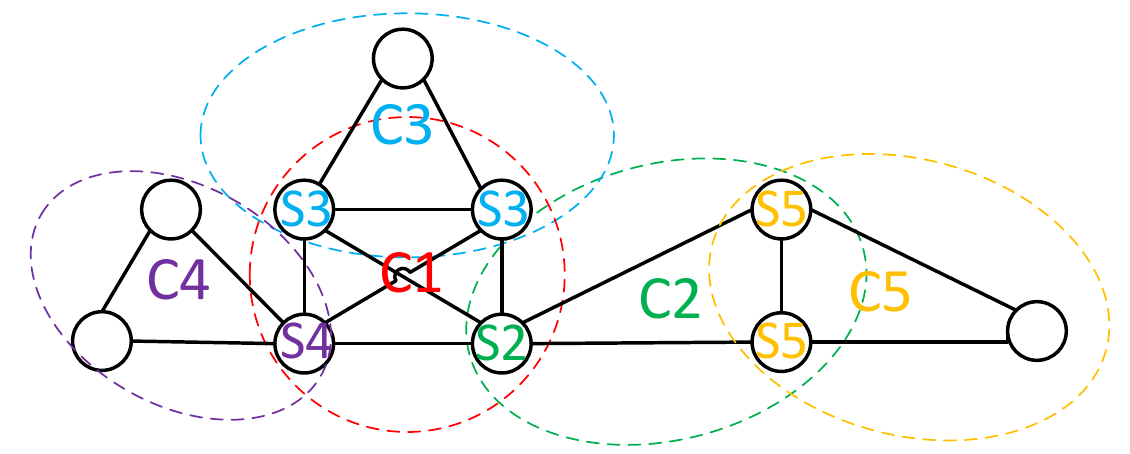}
\caption{The decomposable Gaussian graphical model with 5 cliques and numbered separators.}
\label{conv_fig0}
\end{figure}

Consider the example in Fig. \ref{conv_fig0}, by employing (\ref{Define_q_k}), we observe that $q(2)=q(3)=q(4)=1$ and $q(5)=2$ which implies ${{\cal S}_2} \subseteq {{\cal C}_{1}}$, ${{\cal S}_3} \subseteq {{\cal C}_{1}}$, ${{\cal S}_4} \subseteq {{\cal C}_{1}}$ and ${{\cal S}_5} \subseteq {{\cal C}_{2}}$. By employing (\ref{S_k_C_k1}), we also obtain ${{\cal S}_2} \subseteq {{\cal C}_{2}}$, ${{\cal S}_3} \subseteq {{\cal C}_{3}}$, ${{\cal S}_4} \subseteq {{\cal C}_{4}}$ and ${{\cal S}_5} \subseteq {{\cal C}_{5}}$. Note that $q(2)<2$, $q(3)<3$, $q(4)<4$ and $q(5)<5$. By employing (\ref{Define_Qj}), we obtain ${{\cal Q}_1}=\{2, 3, 4\}$ and ${{\cal Q}_2}=\{5\}$ but ${{\cal Q}_3} = {{\cal Q}_4} = {{\cal Q}_5}= \emptyset$.

\begin{figure}[!t]
\centering
\includegraphics[width=3.5in]{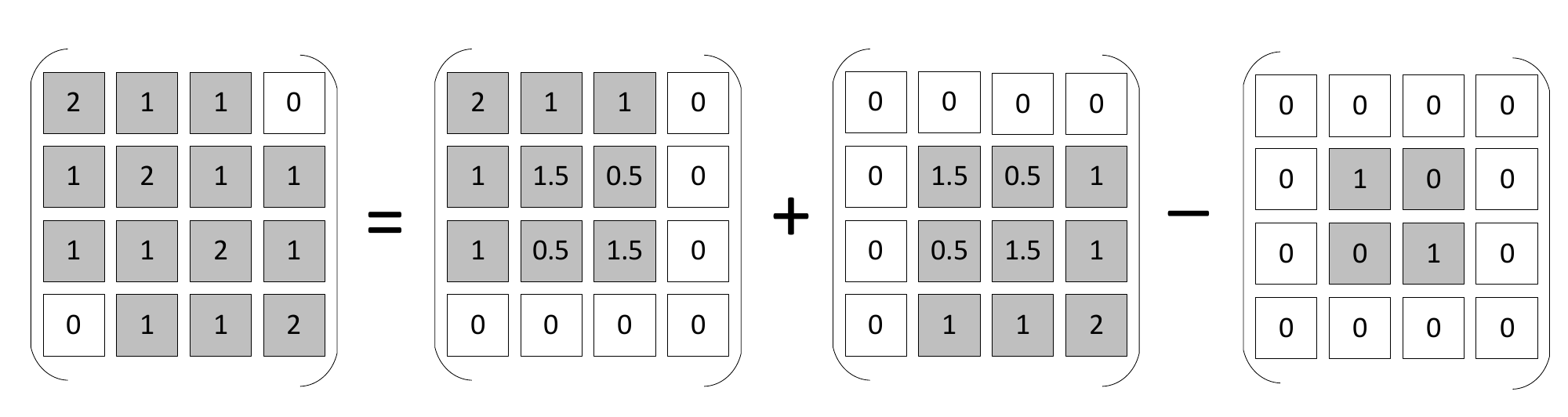}
\caption{A numerical example of equation (\ref{Sigmadivid}) for a two-cluster chain structure (see Fig. \ref{conv_fig1}). Cluster 1: Nodes 1,2,3; Cluster 2: Nodes 2,3,4; Separator: Nodes 2, 3.}
\label{conv_fig10}
\end{figure}

Let ${\bf x}_{{{{\cal C}_k}}}$ denote the set of observations in ${\bf x}$ that come from the nodes in the $k$-th clique. Let ${\bf x}_{{{{\cal S}_k}}}$ denote the observations in ${\bf x}$ that come from the nodes in the $k$-th separator set. Let ${{{\bf{\Sigma }}_{{{\cal C}_k}}}}$ and ${{{\bf{\Sigma }}_{{{\cal S}_k}}}}$ denote the covariance matrices associated with ${\bf x}_{{{{\cal C}_k}}}$ and ${\bf x}_{{{{\cal S}_k}}}$ respectively. The information matrix of $\textbf{x}$ can be expressed by those of the cliques and separators in decomposable GGMs as \cite{lauritzen1996graphical}
\begin{align}\label{Sigmadivid}
{\bf\Sigma}^{-1} = {\sum\limits_{k = 1}^K {{{[({\bf\Sigma}_{{\cal C}_k})^{-1}]}^{\cal V}}} }-{\sum\limits_{k = 2}^K {{{[({\bf\Sigma}_{{\cal S}_k})^{-1}]}^{\cal V}}} }.
\end{align}
Fig. \ref{conv_fig10} gives a numerical example of (\ref{Sigmadivid}), using a chain structure (see Fig. \ref{conv_fig1}). Here, there are $K=2$ clusters, and each cluster has 3 nodes, resulting in 3 terms on the right-hand side of (\ref{Sigmadivid}).

Also, for a decomposable GGM \cite{lauritzen1996graphical},
\begin{align}\label{determinant}
\det{\bf\Sigma} = \Big(\prod\limits_{k = 1}^K \det{\bf\Sigma}_{{\cal C}_k}\Big)\bigg/\Big({{\prod\limits_{k = 2}^K \det{\bf\Sigma}_{{\cal S}_k}}}\Big).
\end{align}

\subsection{Gaussian Covariance Matrix Testing Problems}
Consider a network with $N$ sensors and the Gaussian covariance matrix testing problem given by
\begin{align}\label{general_signal model}
&H_0:\textbf{x} \sim \mathcal{N}(0,\bf{I})\nonumber\\
&H_1:\textbf{x} \sim \mathcal{N}(0,\bf{\Sigma})
\end{align}
where the covariance matrix $\bf \Sigma$ is assumed to be positive definite and without loss of generality we assume the identity matrix under $H_0$, because (\ref{general_signal model}) can be obtained via pre-whitening.

The Bayes optimum decision rule for (\ref{general_signal model}) is a log-likelihood ratio (LLR) threshold test
\begin{equation} \label{Bayes_testing}
{\delta _B}({\bf{x}}) = \left\{ \begin{array}{l}
1 \; \text{ if } T({\bf x}) \ge  2\tau \\
0 \; \text{ if }  T({\bf x}) <  2\tau
\end{array} \right.,
\end{equation}
and the threshold $\tau$ is
\begin{equation} \label{orignalThre}
\tau \buildrel \Delta \over = \ln{{\pi_0}/{\pi_1}}
\end{equation}
where ${\pi_0}$ and ${\pi_1}$ are the a priori probabilities of $H_0$ and $H_1$ respectively. The detector decides $H_1$ if
\begin{align}\label{LLRcenter}
T({\bf x}) &=  2\ln \frac{{f({\bf x}|{H_1})}}{{f({\bf x}|{H_0})}}\nonumber\\
 &= 2\ln \frac{{{{(2\pi )}^{ - \frac{N}{2}}}{{(\det {\bf \Sigma })}^{ - \frac{1}{2}}}\exp \{  - \frac{1}{2}{\bf {x}^T}{{({\bf \Sigma})}^{ - 1}}\bf {x}\} }}{{{{(2\pi )}^{ - \frac{N}{2}}}{{(\det {\bf I})}^{ - \frac{1}{2}}}\exp \{  - \frac{1}{2}{\bf x^T}{{({\bf I})}^{ - 1}}\bf x\} }}\nonumber\\
 &= {\bf {x}^T}{{\bf I}^{ - 1}}{\bf {x}} - {\bf x^T}{{\bf \Sigma }^{ - 1}}{\bf {x}}\nonumber\\
  &\qquad  - \ln \det ({\bf \Sigma}) + \ln \det ({\bf I})  > 2\tau.
\end{align}

The optimum centralized energy unconstrained detection approach requires
each sensor to send its observation to a FC. After receiving the
data from all sensors, the FC employs (\ref{LLRcenter}).
In our distributed approach, we partition the sensors into $K$ clusters which correspond to the cliques. The CH will collect the information from the sensors in a given cluster and then transmit it to the FC. By employing ordering over the clusters, we reduce the number of CH to FC transmissions while achieving the same probability of error as the optimum centralized energy unconstrained detection approach. Throughout the paper, the following assumption is made.

\begin{assumption} \label{Assumption_local_communication}
	Every sensor is physically close to its neighboring sensors in the graph. Hence, sensors in the same cluster are physically close so that intercluster communications used by CHs are assumed to be short distances so we only focus on communications between the CHs and the FC.
\end{assumption}

Plugging (\ref{Sigmadivid}) and (\ref{determinant}) into (\ref{LLRcenter}), the test statistics $T(\bf x)$ can be rewritten as (refer to Appendix A for details)
\begin{align}
T({\bf x})= \sum_{k=1}^K \big({{\bf {x}}^T_{{{\cal C}_K}}} {{ {{{\bf{J}}_k}} }} {{\bf {x}}_{{{\cal C}_K}}} -{ e}_k \big)
 = \sum\limits_{k = 1}^K {{L_k}({{\bf{x}}_{{{\cal C}_k}}})}\label{sum_LLLR_local22}
\end{align}
where
\begin{align}\label{J1}
\!\!{{{\bf{J}}_1}}& =  \Big({{\bf I}_{{{\cal C}_1}}^{-1}}- ({\bf{\Sigma}}_{{{\cal C}_1}})^{-1}\Big) -  \sum\limits_{j \in {{\cal Q}_1}}{\beta _j} \Big[ {{\bf I}_{{{\cal S}_j}}^{-1}} - ({\bf{\Sigma}}_{{{\cal S}_j}})^{-1} \Big]^{{{\cal C}_1}},
\end{align}
and
\begin{align}
{{{\bf{J}}_k}}& = \Big({{\bf I}_{{{\cal C}_k}}^{-1}}- ({\bf{\Sigma}}_{{{\cal C}_k}})^{-1}\Big) - \sum\limits_{j \in {{\cal Q}_k}}{\beta _j}\Big[ {{\bf I}_{{{\cal S}_{j}}}^{-1}} - ({\bf{\Sigma}}_{{{\cal S}_{j}}})^{-1} \Big]^{{{\cal C}_k}} \notag \\
& \qquad - \alpha_k \Big[ {{\bf I}_{{{\cal S}_{k}}}^{-1}} - ({\bf{\Sigma}}_{{{\cal S}_{k}}})^{-1} \Big]^{{{\cal C}_k}}\label{Jk}
\end{align}
for $k=2,3,...,K$ where ${\cal Q}_j$ for $j=1,2,...,K$ is defined in (\ref{Define_Qj}) and the set of coefficient pairs $\{(\alpha_k, \beta_k)\}_{k=2}^K$ satisfies
\begin{equation} \label{alpha_beta_1}
{\alpha _k} + {\beta _k} = 1, \; \forall k=2,3,...,K.
\end{equation}
Note that there are uncountably many choices of $\{\alpha_k,\beta_k\}_{k=2}^K$ satisfying (\ref{alpha_beta_1}) which implies that multiple $T({\bf x})$ in (\ref{sum_LLLR_local22}) will work where each splits the observations a little differently over the clusters via nonuniqueness of separator sets.
In (\ref{sum_LLLR_local22}),
\begin{align}
{e}_1 = \ln\det({\bf{\Sigma}}_{{{\cal C}_1}})- \sum\limits_{j \in {{\cal Q}_1}}{\beta _j} \ln\det({\bf{\Sigma}}_{{{\cal S}_j}})\label{e1},
\end{align}
and for all $k=2,3,...,K$,
\begin{align}
\!\!\!\!{e}_k &= \!\!\ln\det({\bf{\Sigma}}_{{{\cal C}_k}})\!-\!\! \!\! \sum\limits_{j \in {{\cal Q}_k}}{\beta _j} \ln\det({\bf{\Sigma}}_{{{\cal S}_{j}}})\!-\!\alpha_k\ln\det({\bf{\Sigma}}_{{{\cal S}_{k}}}).\label{Mk}
\end{align}
We refer to the $k$-th local test statistic (LTS) ${{L_k}({{\bf{x}}_{{{\cal C}_k}}})}$ as the local log-likelihood ratio (LLLR), since each ${L_k}\left( {{{\bf{x}}_{{{\cal C}_k}}}} \right)$ can be calculated by only utilizing the observations available in the $k$-th cluster for $k=1,2,...,K$.

%

\section{Reduced Transmission Energy Detection Using Ordered Transmission Approach}\label{reducedtrans}
\label{Section_Ordering}


\subsection{Ordering for the Covariance Testing Problem}\label{Orderingforcovariance}
The following theorem describes our distributed algorithm (ordering) to reduce transmissions \cite{4545248}.
\begin{theorem} \label{Theorem_Ordering}
Consider an approach where the $k$-th CH will transmit its ${{L_k}({{\bf{x}}_{{{\cal C}_k}}})}$ from (\ref{sum_LLLR_local22}) after a time equal to $\eta /|{{L_k}( {{{\bf{x}}_{{{\cal C}_k}}}} )}|$, with $\eta$ any real positive number. All transmissions stop when the sum of the ${{L_k}({{\bf{x}}_{{{\cal C}_k}}})}$ received is larger than a threshold $\tau_U$ or smaller than a threshold $\tau_L$. The ordered magnitudes of $\{L_k({\bf x}_{{\cal C}_k})\}_{k=1}^K$ are denoted as
\begin{equation}\label{orderLLRATIO}
|\tilde L_{1} | >
|\tilde L_{2}| > \cdots
|\tilde L_{K} |.
\end{equation}
Let $n_{UT}<K$ be the number of CHs who have not yet transmitted at a given time and let ${\tilde L}_{K-n_{UT}}$ denote the last LLLR transmission. Define
\begin{equation} \label{Define_tau_U}
\tau_U \buildrel \Delta \over = 2\tau + n_{UT}|{\tilde L}_{K-n_{UT}}|
\end{equation}
and
\begin{equation} \label{Define_tau_L}
\tau_L \buildrel \Delta \over = 2\tau - n_{UT}|{\tilde L}_{K-n_{UT}}|,
\end{equation}
where $\tau$ is defined in (\ref{orignalThre}). The approach described in this theorem gives the same probability of error as the optimum centralized energy unconstrained approach where all CHs transmit their LLLRs to the FC, while using a smaller average number of transmissions.
\end{theorem}
\begin{IEEEproof}
	The proof of this theorem is omitted, since it follows the proof in \cite{4545248}.
\end{IEEEproof}

As demonstrated by \emph{Theorem \ref{Theorem_Ordering}}, every time the FC receives a new CH transmission, it updates the thresholds and compares them with the sum of the CH transmissions it has received so far, denoted by $L'$. If $L' \ge {\tau}_U$ (or $L' < {\tau}_L$), then the test statistic $T({\bf x})$ from (\ref{LLRcenter}) is guaranteed to be larger (or smaller) than the optimum threshold $2\tau$, regardless of the CH transmissions that have not yet been transmitted. Then the FC can send a message to the CHs to stop the remaining transmissions. By halting before all transmissions have taken place, transmissions can be saved without performance loss when sufficient evidence is accumulated. If ${\tau}_L \le L' < {\tau}_U$, which means the FC has not accumulated sufficient evidence yet, then transmissions continue. Note that the ordering approach is only applied to the global communications (communications from the CHs to the FC), but not local communications (intercluster communications).

\subsection{Lower Bound on the Average Number of Transmissions Saved}
For any given  $\delta$, we define the detection probability $P_{D,k}$ and the false alarm probability $P_{f,k}$ in the $k$-th cluster for $k=1,2,...,K$ as
\begin{equation} \label{Pr_H1_LB_Lemma2}
P_{D,k}(\delta) = \Pr \left( {{L_k}\left( {{{\bf{x}}_{{{\cal C}_k}}}} \right)  > \delta \left| {{H_1}}  \right.} \right)
\end{equation}
and
\begin{equation} \label{Pr_H0_LB_Lemma2}
	P_{f,k}(\delta) = 1- \Pr \left( {{L_k}\left( {{{\bf{x}}_{{{\cal C}_k}}}} \right)  \le  \delta \left| {{H_0}}  \right.} \right).
\end{equation}

We provide the following theorem with regard to a lower bound on the average number of transmissions saved by ordering the global communications.
\begin{theorem} \label{Theorem_saving}
Consider the hypothesis testing problem described in (\ref{general_signal model}).
Let
\begin{equation} \label{Delta_0}
{\delta ^{\left( 0 \right)}} \buildrel \Delta \over = \min \left\{ {2\tau ,0} \right\},
\end{equation}
and
\begin{equation} \label{Delta_1}
{\delta ^{\left( 1 \right)}} \buildrel \Delta \over = \max \left\{ {2\tau ,0} \right\}
\end{equation}
where the threshold $\tau$ is defined in (\ref{orignalThre}).
When using the ordered transmission approach described in \emph{Theorem \ref{Theorem_Ordering}},
if $K > 1$,
then for any choice of $\{\alpha_k, \beta_k\}_{k=2}^K$, the average number of transmissions saved $K_s$ by ordering the global communications is bounded from below by
\begin{align} \notag
{K_s} > & \max\Bigg\{0, \left( {\left\lceil {\frac{K}{2}} \right\rceil  - 1} \right){\pi _1}{\sum\limits_{k = 1}^K {{{ P_{D,k}({\delta ^{\left( 1 \right)}})  } } } } \\  \notag
& \quad \; + \left( {\left\lceil {\frac{K}{2}} \right\rceil  - 1} \right){\pi _0}\sum\limits_{k = 1}^K \big(1-P_{f,k}({\delta ^{\left( 0 \right)}})\big) \\ \label{Theorem_state_K_s}
& \quad \;  - \left( {\left\lceil {\frac{K}{2}} \right\rceil  - 1} \right)\left( {K - 1} \right)\Bigg\},
\end{align}
where ${ P_{D,k}({\delta ^{\left( 1 \right)}})  }$ and $P_{f,k}({\delta ^{\left( 0 \right)}})$ were previously defined in (\ref{Pr_H1_LB_Lemma2}) and (\ref{Pr_H0_LB_Lemma2}), respectively.
\end{theorem}
\begin{IEEEproof}
	The proof of this theorem is omitted, since it follows the proof in \cite{zhang2017ordering}. The difference from \cite{zhang2017ordering} is that we need to employ ${ P_{D,k}({\delta ^{\left( 1 \right)}})  }$ in (\ref{Pr_H1_LB_Lemma2}) and $P_{f,k}({\delta ^{\left( 0 \right)}})$ in (\ref{Pr_H0_LB_Lemma2}) which are completely different for this new problem. The other difference from \cite{zhang2017ordering} is that we add zero as a choice in (\ref{Theorem_state_K_s}) since the average number of transmissions saved $K_s$ makes sense when it is non-negative in reality.
\end{IEEEproof}

The lower bound on the average number of transmissions saved by ordering the global communications as demonstrated in \emph{Theorem \ref{Theorem_saving}} is  valid for any choice of $\{\alpha_k, \beta_k\}_{k=2}^K$ and any positive definite matrix $\bf \Sigma$ corresponding to a decomposable GGM. According to the expression on the right-hand side of (\ref{Theorem_state_K_s}), the lower bound on the average number of transmissions saved by ordering the global communications depends on the entries of the covariance matrix ${\bf \Sigma}$, the parameters $\{\alpha_k, \beta_k\}_{k=2}^K$ and the prior probabilities $\pi_0$ and $\pi_1$.

Define
\begin{equation} \label{mineigallclusters}
\lambda_{\min}=\min_{k}\lambda_{\min,k}
\end{equation}
where $\lambda_{\min,k}$ is the minimum eigenvalue of ${\bf{\Sigma}}_{{{\cal C}_k}}$ and $k=1,2,...,K$.
The following theorem describes a lower bound on the average number of transmissions saved by ordering the global communications when $\lambda_{\min}$ is large.

%

\begin{theorem}\label{limitingbehavior}
Consider the ordered transmission approach described in \emph{Theorem \ref{Theorem_Ordering}} which employs (\ref{sum_LLLR_local22}) with

\begin{equation} \label{alpha_Theorem2}
{\alpha _k} = 1 - {2^{K - k}}\gamma,
\end{equation}
and
\begin{equation}  \label{beta_Theorem2}
{\beta _k} = {2^{K - k}}\gamma,
\end{equation}
for all  $k=2,3,...,K$ using any $\gamma $ which satisfies
\begin{equation} \label{gamma_Theorem2}
\gamma  \in \left( {0,\frac{1}{{{2^{K - 1}} - 1}}} \right).
\end{equation}
With a sufficiently large eigenvalue $\lambda_{\min}$ defined in (\ref{mineigallclusters}) and $K > 1$,
the average number of transmissions saved $K_s$ by ordering the global communications is bounded from below by
\begin{equation} \label{Corollary_state_K_s}
K_s > {\left\lceil {\frac{K}{2}} \right\rceil  - 1}.
\end{equation}

\end{theorem}

\begin{IEEEproof}
Refer to Appendix B.
\end{IEEEproof}
Note that \emph{Theorem \ref{limitingbehavior}} states that the average number of transmissions saved by ordering the global communications increases
at least as fast as linearly proportional to the number of clusters $K$
while achieving the same probability of error as the conventional centralized
detection approach. There are uncountable choices of $\{(\alpha_k, \beta_k)\}_{k=2}^K$ to guarantee the lower bound where the average number of transmissions saved can be larger than half the number of clusters employed.

The minimum eigenvalue of ${\bf{\Sigma}}_{{{\cal C}_k}}$, denoted $\lambda_{\min,k}$, approaching infinity leads the Kullback-Leibler divergence  to increase to infinity.  Some other statistical distance measures which quantify the distance between two distributions can also be employed here and lead to the same conclusion, for example the Bhattacharyya distance. The Kullback-Leibler divergence ${D_{{\text{KL}}}}( {f( {{{\bf{x}}_{{{\cal C}_k}}}| {{H_0}} } )\| {f( {{{\bf{x}}_{{{\cal C}_k}}}| {{H_1}} } )} } )$ between $f( {{{\bf{x}}_{{{\cal C}_k}}}| {{H_0}} } )$ and $f( {{{\bf{x}}_{{{\cal C}_k}}}| {{H_1}} } )$ increases to infinity for all $k$, since
\begin{align} \notag
& \mathop {\lim }\limits_{\lambda_{\min,k} \to \infty } {D_{{\text{KL}}}}\left( {f\left( {{{\bf{x}}_{{{\cal C}_k}}}\left| {{H_0}} \right.} \right)\left\| {f\left( {{{\bf{x}}_{{{\cal C}_k}}}\left| {{H_1}} \right.} \right)} \right.} \right)\\  \notag
& = \mathop {\lim }\limits_{\lambda_{\min,k} \to \infty} {{\mathbbm{E}}_{f\left( {{{\bf{x}}_{{{\cal C}_k}}}\left| {{H_0}} \right.} \right)}}\left\{ {\ln \frac{{f\left( {{{\bf{x}}_{{{\cal C}_k}}}\left| {{H_0}} \right.} \right)}}{{f\left( {{{\bf{x}}_{{{\cal C}_k}}}\left| {{H_1}} \right.} \right)}}} \right\}\\  \notag
& =  \mathop {\lim }\limits_{\lambda_{\min,k} \to \infty }\frac{1}{2}\bigg[ log\det{\bm \Sigma}_{{{\cal C}_k}} + trac({\bm \Sigma}_{{{\cal C}_k}}^{-1} ) - M_k \bigg]   \\  \notag
& = \mathop {\lim }\limits_{\lambda_{\min,k} \to \infty } \frac{1}{2} \bigg[  \sum_{m=0}^{M_k-1} \big( \log\lambda_{m,k} + \frac{1}{\lambda_{m,k}} \big)  - M_k \bigg] \\  \notag
& \rightarrow +\infty.
\end{align}
where $M_k$ is the dimension of ${\bm \Sigma}_{{{\cal C}_k}}$ and $\lambda_{m,k}$ is the $m$-th eigenvalue of ${\bm \Sigma}_{{{\cal C}_k}}$. Thus, the fact that the significant average number of transmissions saved can be obtained as shown in \emph{Theorem \ref{limitingbehavior}} is consistent with our intuition that with a large distance between the cluster covariance matrices under two hypotheses, they are easy to tell apart so we do not need many observations to make a reliable decision. Note that $\lambda_{\min,k}\rightarrow\infty$ for all $k=1,2,...,K$ is only a sufficient condition rather than a necessary condition. Deriving necessary and sufficient conditions for saving half of the transmissions for the covariance matrix testing problem will be pursued in our future work.

\section{Numerical Results}\label{numericalresu}

In this section, numerical examples of two representative classes
of decomposable Gaussian graphical models with chain structure and tree structure are presented to illustrate the lower bounds on the number of transmissions saved by employing ordered transmissions.

\subsection{Average Percentage of Transmissions Saved versus the Minimum Eigenvalue}\label{Average_Percentage_Section}

In this subsection, the theoretical lower bound in (\ref{Theorem_state_K_s}) is compared with the average number of transmissions saved by ordering the CH to FC communications from Monte Carlo simulations. Consider a class of decomposable Gaussian graphical models with $20$ clusters as shown in Fig. \ref{conv_fig1}. As depicted in Fig. \ref{conv_fig1}, each cluster consists of $5$ sensors, and every two consecutive clusters are coupled through a $1$-sensor separator. In the simulation results in Fig. \ref{conv_fig2}, we set $\pi_0=\pi_1=0.5$ and $\gamma = 0.5/(2^{19} - 1)$. Assume the eigenvalues of ${\bf\Sigma}_{{\cal C}_1}$ form a vector with five elements which are uniformly distributed from $\alpha$ to $1.5\alpha$ where $\alpha$ can be varied to change the minimum eigenvalue of ${\bf\Sigma}_{{\cal C}_1}$. We generate a diagonal matrix ${\bf \Lambda}$ with these eigenvalues along the diagonal and use a unitary matrix ${\bf V}$ to construct a non-diagonal matrix ${\bf\Sigma}_{{\cal C}_1}={\bf V}^T {\bf \Lambda} {\bf V}$ where the unitary matrix is generated by employing Gram–-Schmidt orthonormalization \cite{horn_johnson_2012} while ensuring (\ref{S_k_C_k}) and (\ref{S_k_C_k1}) are satisfied. We take ${\bf\Sigma}_{{\cal C}_k}$ to be equal to ${\bf\Sigma}_{{\cal C}_1}$ for $k=2,3,...,K$. We use the same ${\bf\Sigma}_{{\cal C}_k}$ during all the Monte Carlo runs so the statistical descriptions of the observations are the same.  However, the actual observations are randomly regenerated in each Monte Carlo run so they are different over each Monte Carlo run. We employ the Monte Carlo method (20000 runs) to obtain the average number of transmissions saved by ordering the global communications as illustrated in Fig. \ref{conv_fig2} for different values of $\alpha$. For comparison, in Fig. \ref{conv_fig2}, the theoretical lower bound in (\ref{Theorem_state_K_s}) is also provided. It is shown that as the parameter $\alpha$ increases (the minimum eigenvalue $\lambda_{\min}$ increases), the value of the theoretical lower bound on the average number of transmissions saved by ordering the global communications increases. As expected from our analysis, the simulation results in Fig. \ref{conv_fig2} show that the lower bound on the average percentage of transmissions saved nearly equals to 0.45 when $a=199$ which is consistent with \emph{Theorem \ref{limitingbehavior}} since ${(\left\lceil {\frac{K}{2}} \right\rceil  - 1)/K}=0.45$ when $K=20$.  These results also illustrate that the theoretical lower
bound in \emph{Theorem \ref{Theorem_saving}} is also a valid lower bound on the simulated estimates for the average number of transmissions saved by employing ordered transmissions for the specific cases considered.

\subsection{Average Number of Transmissions Saved versus the Number of Clusters}

\begin{figure}[!t]
\centering
\includegraphics[width=3.5in]{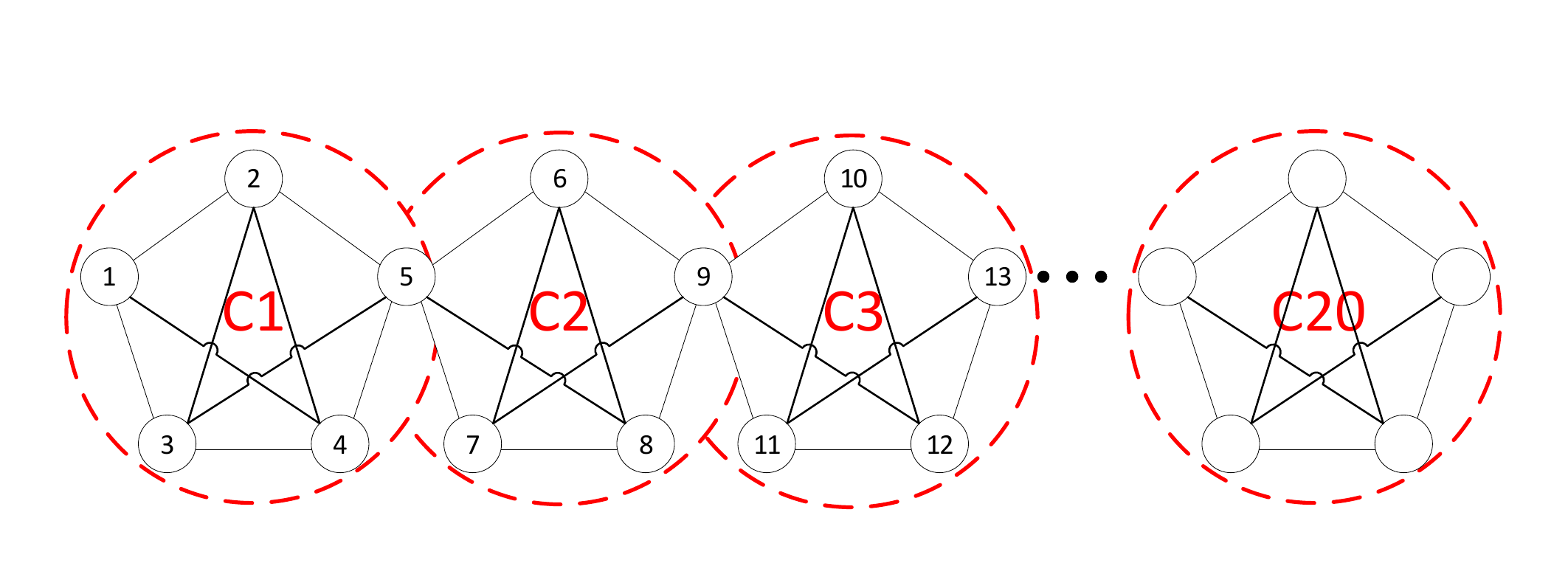}
\caption{The decomposable Gaussian graphical model with chain structure. Clique 1: Nodes 1,2,3,4,5; Separator: Node 5; Clique 2: Nodes 5,6,7,8,9; Separator: Node 9; the rest is similar.}
\label{conv_fig1}
\end{figure}

\begin{figure}[!t]
\centering
\includegraphics[width=3.5in]{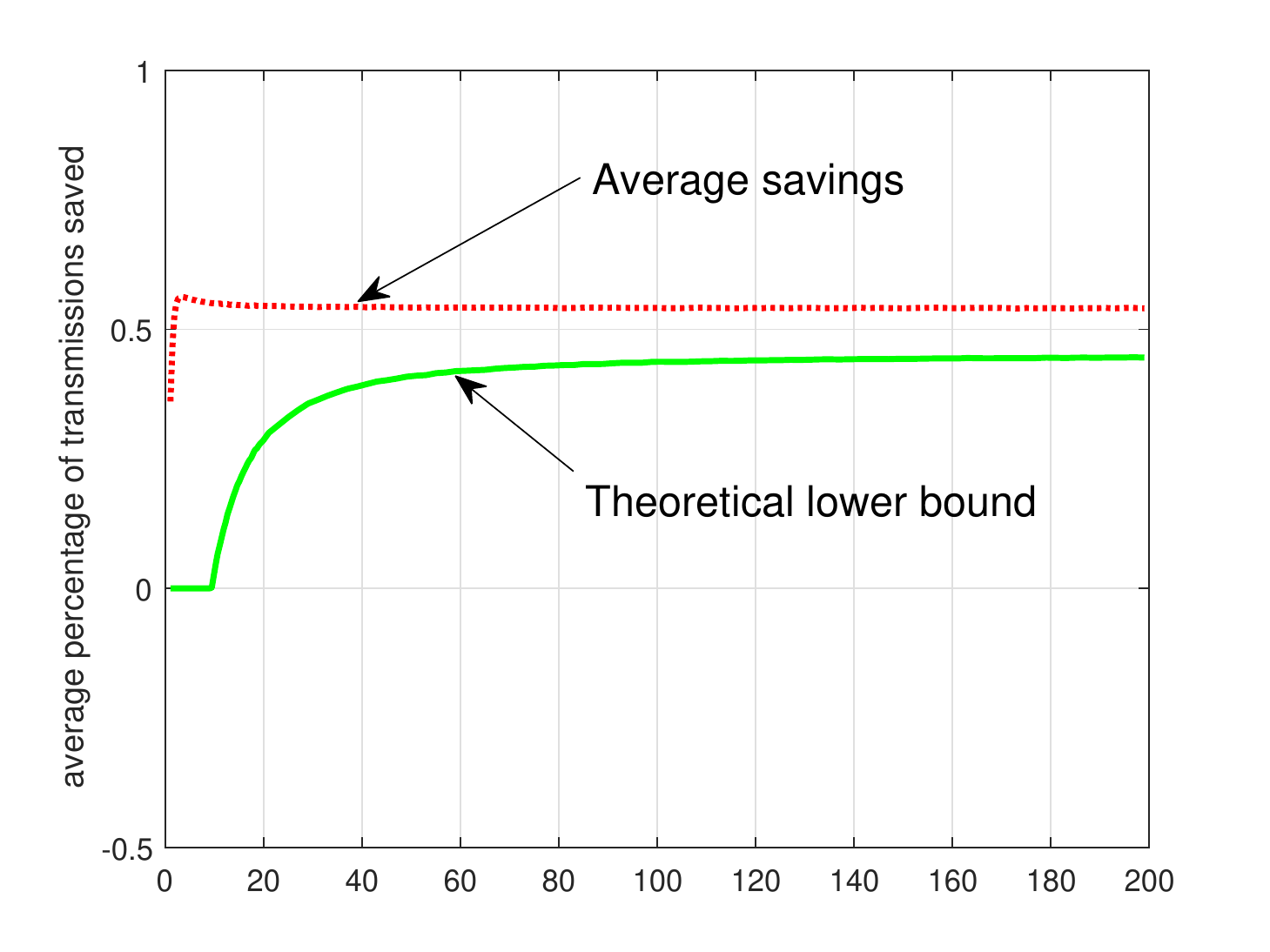}
\caption{Impact of the parameter $a$ on the average percentage of transmissions saved for the model illustrated in Fig. \ref{conv_fig1}. }
\label{conv_fig2}
\end{figure}

\begin{figure}[!t]
\centering
\includegraphics[width=3.5in]{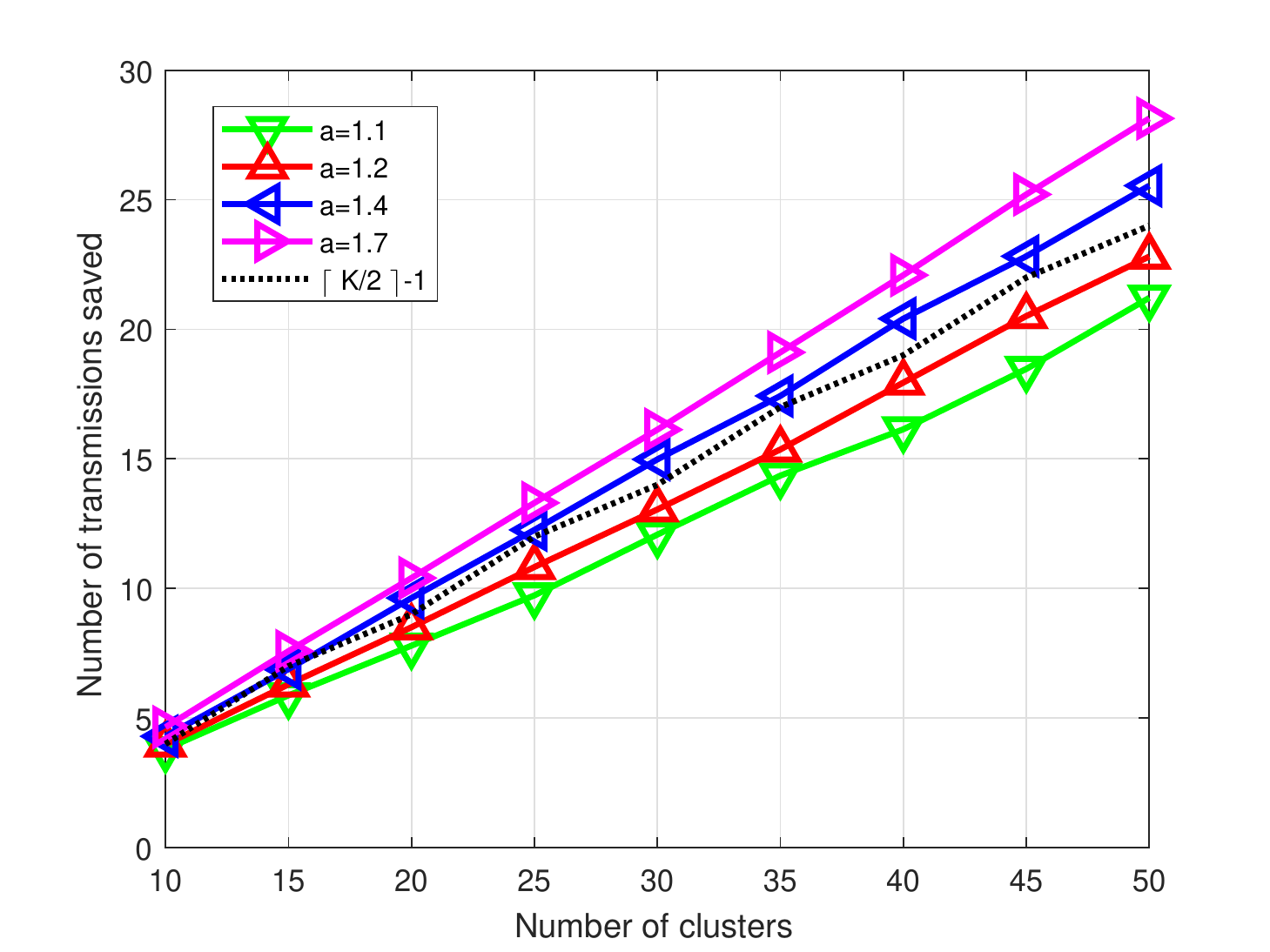}
\caption{The average number of transmissions saved versus $K$ for the model illustrated in Fig. \ref{conv_fig1}. }
\label{conv_fig3}
\end{figure}

In this subsection, we investigate the average number of transmissions saved by ordering the global communications for a different number of clusters $K$. We consider the same class of decomposable Gaussian graphical models as in Fig. \ref{conv_fig1}. In the simulation, we also set $\pi_0=\pi_1=0.5$ and $\gamma = 0.5/(2^{K} - 1)$ and generate ${\bf\Sigma}_{{\cal C}_k}$ for $k=1,2,...,K$ in the same way as shown in the previous subsection. Fig. \ref{conv_fig3} illustrates the average number of transmissions saved by ordering the global communications versus $K$ for different values of $\alpha$.  For comparison, in Fig. \ref{conv_fig3}, the limiting theoretical lower bound in (\ref{Corollary_state_K_s}) is also provided. Fig. \ref{conv_fig3} illustrates that the average number of transmissions saved by ordering the global communications  increases approximately linearly with $K$ for
every value of $\alpha$. It indicates that the rate of increase with $K$ becomes faster when the parameter $\alpha$ is increased. In addition, for the sufficiently large minimum eigenvalue $\alpha>1.4$ in this scenario, ${\left\lceil {\frac{K}{2}} \right\rceil  - 1}$ asymptotically serves as a lower bound on the number of transmissions saved by ordering the global communications.

\begin{figure}[!t]
\centering
\includegraphics[width=3.5in]{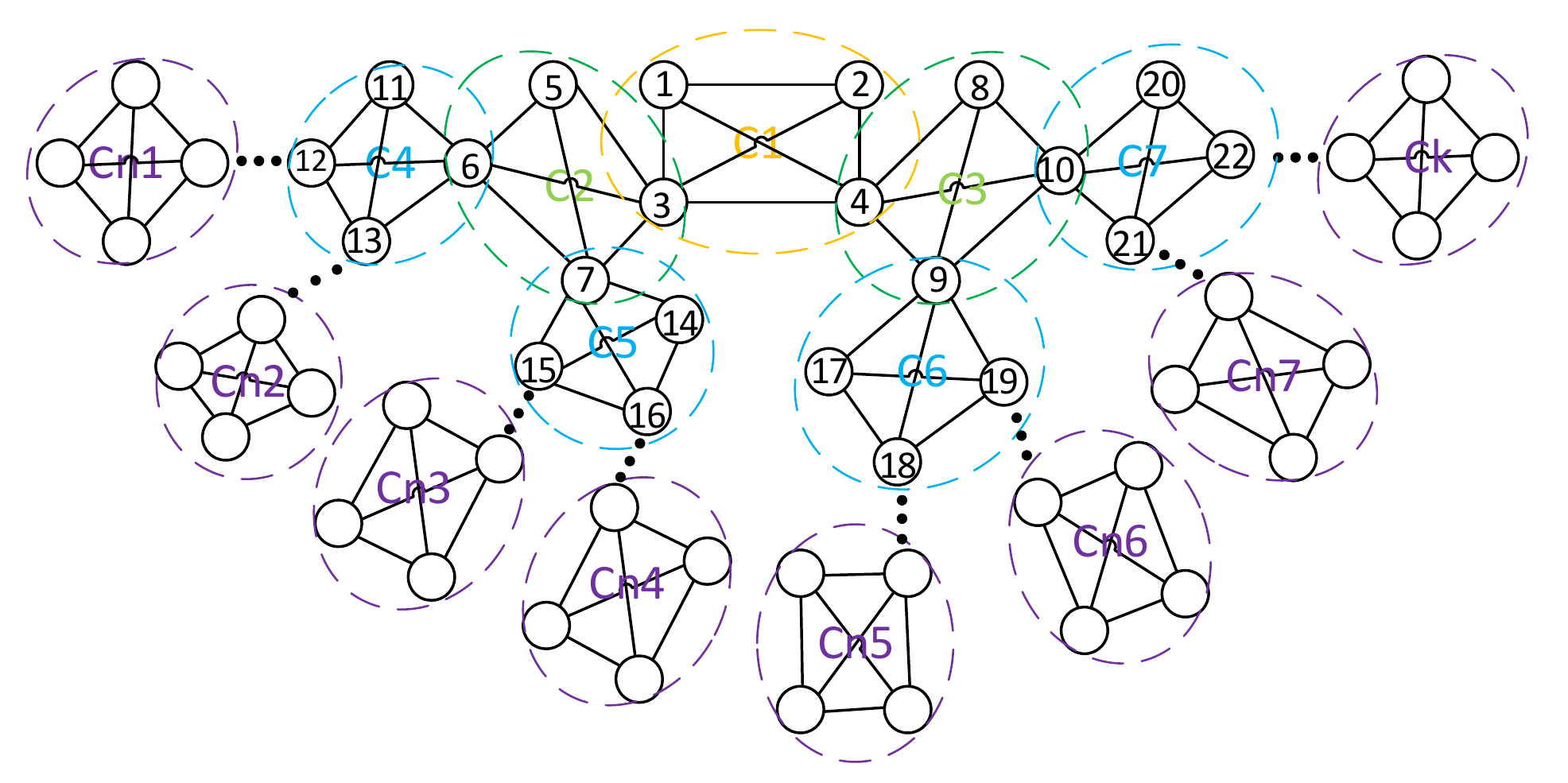}
\caption{The decomposable Gaussian graphical model with binary tree structure. Clique 1: Nodes 1,2,3,4; Separator: Nodes 3,4; Clique 2: Nodes 3,5,6,7; Separator: Nodes 3,6,7; the rest is similar.}
\label{conv_fig4}
\end{figure}

\begin{figure}[!t]
\centering
\includegraphics[width=3.5in]{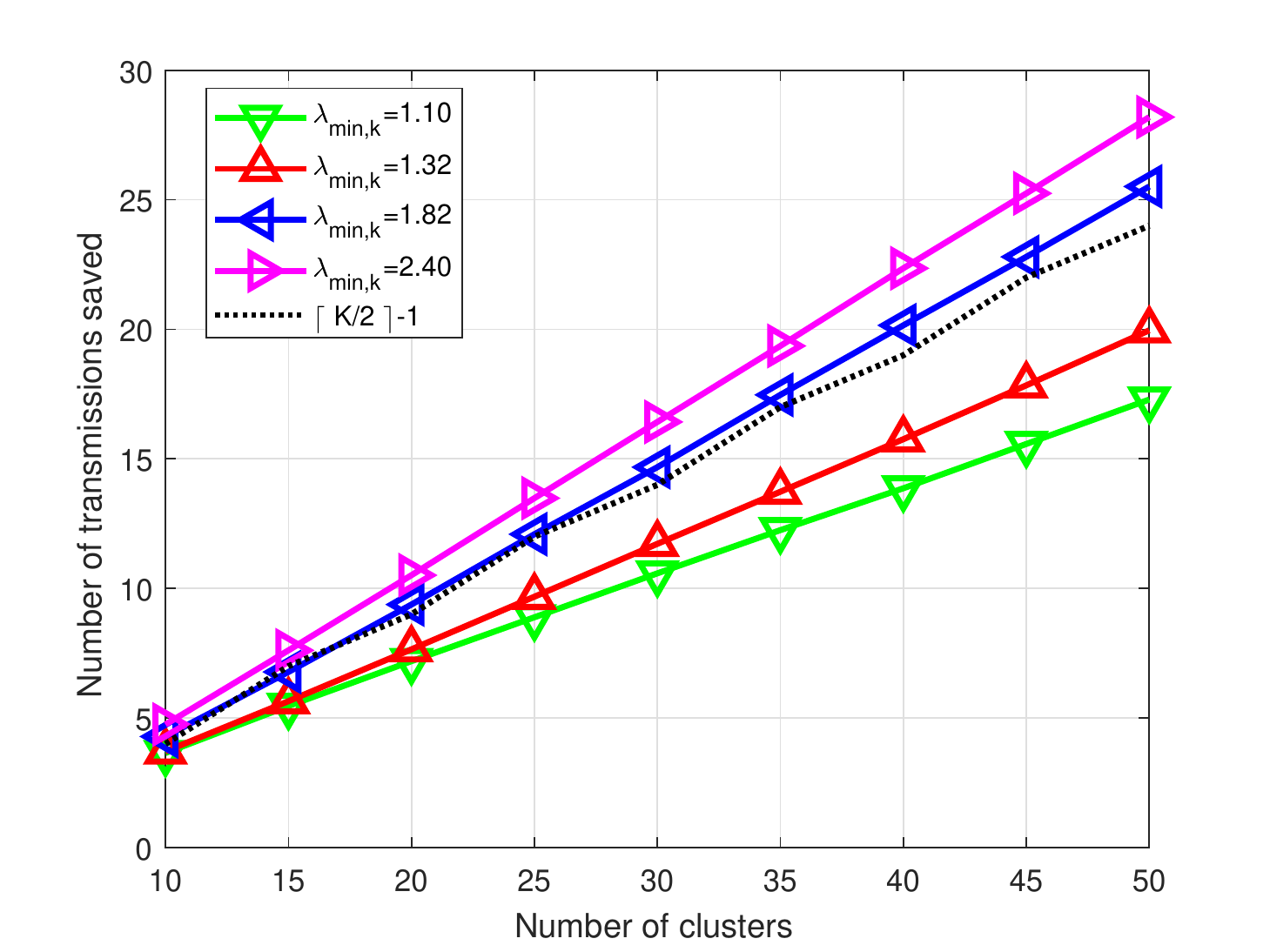}
\caption{The average number of transmissions saved versus $K$ for the model illustrated in Fig. \ref{conv_fig4}.}
\label{conv_fig5}
\end{figure}

Next, we consider a different class of decomposable Gaussian graphical model with the binary tree structure illustrated in Fig. \ref{conv_fig4} where each cluster has 4 nodes and each separator set has 1 node. We set $\pi_0=\pi_1=0.5$ and $\gamma = 0.5/(2^{K} - 1)$. The diagonal elements of ${\bf\Sigma}_{{\cal C}_k}$ are set to be $x^2$ and the other elements of ${\bf\Sigma}_{{\cal C}_k}$ are set to equal to $x/10$ where we change the minimum eigenvalue of ${\bf\Sigma}_{{\cal C}_k}$ by changing the parameter $x$ since the minimum eigenvalue of ${\bf\Sigma}_{{\cal C}_k}$ equals to $x^2-x/10$. We fix each cluster to have the same value of $x$ in each iteration. We consider the average number of transmissions saved versus the number of clusters for $x=1.1, 1.2, 1.4$ and $1.6$ which correspond to $\lambda_{\min,k}=1.10, 1.32, 1.82$ and $2.40$, respectively.  Fig. \ref{conv_fig5} implies the average number of transmissions saved by ordering the
global communications increases approximately linearly with $K$ for every value of $\lambda_{\min,k}$. The larger the value of $\lambda_{\min,k}$, the larger the slope of curves which is very similar to the result in Fig. \ref{conv_fig3}.

\section{Conclusion}\label{conclusion}

We have proposed the ordered transmission approach to testing the covariance matrix of a GGM where transmissions can be saved without performance loss. After each CH collects and summarizes the data from the sensor nodes, this approach is employed to reduce the number of transmissions from the CHs to the FC. A lower bound on the average number of transmissions was derived which can approach
approximately half the number of clusters when the minimum
eigenvalue of the covariance matrix under the alternative hypothesis in each cluster becomes
sufficiently large.

\appendices

\section{Details of (\ref{sum_LLLR_local22}) }
Plugging (\ref{Sigmadivid}) and (\ref{determinant}) into (\ref{LLRcenter}), the test statistics $T(\bf x)$ can be rewritten as
\begin{align}
&T({\bf x}) ={\bf {x}^T}\Bigg[ \Big( \sum_{k=1}^K[({{\bf I}_{{{\cal C}_k}}})^{-1}]^{\cal V} - \sum_{k=2}^K[({{\bf I}_{{{\cal S}_k}}})^{-1}]^{\cal V}\Big)\\  \notag
&\qquad - \Big(\sum_{k=1}^K[{\bf{\Sigma}}_{{{\cal C}_k}}^{-1}]^{\cal V}  -  \sum_{k=2}^K[{\bf{\Sigma}}_{{{\cal S}_k}}^{-1}]^{\cal V}\Big) \Bigg] {\bf {x}}\\
&\qquad-\sum_{k=1}^K \ln \det({\bf{\Sigma}}_{{{\cal C}_k}}) + \sum_{k=2}^K \ln \det({\bf{\Sigma}}_{{{\cal S}_k}})\label{dividecluster}\\ \notag
& =  {\bf {x}^T} \Bigg[ \Big[{{\bf I}_{{{\cal C}_1}}^{-1}}- ({\bf{\Sigma}}_{{{\cal C}_1}})^{-1}\Big]^{\cal V} - \sum\limits_{j \in {{\cal Q}_1}}{\beta _j} \Big[ {{\bf I}_{{{\cal S}_j}}^{-1}} - ({\bf{\Sigma}}_{{{\cal S}_j}})^{-1} \Big]^{\cal V}\\ \notag
& \quad + \sum_{k=2}^{K} \Big\{  \Big[{{\bf I}_{{{\cal C}_k}}^{-1}}- ({\bf{\Sigma}}_{{{\cal C}_k}})^{-1}\Big]^{\cal V} - \sum\limits_{j \in {{\cal Q}_k}}{\beta _j}\Big[ {{\bf I}_{{{\cal S}_{j}}}^{-1}} - ({\bf{\Sigma}}_{{{\cal S}_{j}}})^{-1} \Big]^{\cal V}  \\ \notag
& \qquad - \alpha_k \Big[ {{\bf I}_{{{\cal S}_{k}}}^{-1}} - ({\bf{\Sigma}}_{{{\cal S}_{k}}})^{-1} \Big]^{\cal V} \Big\}  \Bigg]{\bf {x}}\\ \notag
&\qquad-\Bigg[\ln\det({\bf{\Sigma}}_{{{\cal C}_1}})- \sum\limits_{j \in {{\cal Q}_1}}{\beta _j} \ln\det({\bf{\Sigma}}_{{{\cal S}_j}}) \\ \notag
&\quad +\sum_{k=2}^{K}\bigg(\ln\det({\bf{\Sigma}}_{{{\cal C}_k}})-  \sum\limits_{j \in {{\cal Q}_k}}{\beta _j} \ln\det({\bf{\Sigma}}_{{{\cal S}_{j}}})\\
&\quad\quad-\alpha_k\ln\det({\bf{\Sigma}}_{{{\cal S}_{k}}})\bigg)\Bigg]\label{partialpart}\\
& =   \sum_{k=1}^K \big( {\bf {x}^T} {{\left[ {{{\bf{J}}_k}} \right]}^{\cal V}} {\bf {x}} -{e}_k \big)\label{wholeseparatevector}  \\
& = \sum_{k=1}^K \big({{\bf {x}}^T_{{{\cal C}_K}}} {{ {{{\bf{J}}_k}} }} {{\bf {x}}_{{{\cal C}_K}}} -{ e}_k \big)\label{separatevector} \\
& = \sum\limits_{k = 1}^K {{L_k}({{\bf{x}}_{{{\cal C}_k}}})}\label{sum_LLLR_local}
\end{align}
In going from (\ref{dividecluster}) to (\ref{partialpart}), we group the separator terms involving the same observations in several ways since the separator set data is actually contained in the data for the associated clusters. Then (\ref{wholeseparatevector}) follows by using the definition in (\ref{J1})--(\ref{Mk}). In going from (\ref{wholeseparatevector}) to (\ref{separatevector}), we only employ the subvector of $\bf x$, denoted as ${\bf {x}}_{{{\cal C}_K}}$, which consists of the elements of $\bf x$ corresponding to the indices contained
in the $k$-th cluster since other subvectors of $\bf x$ will not have an impact for this cluster.

\section{Proof of \emph{Theorem \ref{limitingbehavior} } }
We can bound the eigenvalues of $\big({{\bf I}_{{{\cal C}_k}}^{-1}}- ({\bf{\Sigma}}_{{{\cal C}_k}})^{-1}\big)$ for $k=1,2,...,K$, denoted as $eig\{ {{\bf I}_{{{\cal C}_k}}^{-1}}- ({\bf{\Sigma}}_{{{\cal C}_k}})^{-1} \}$,  as follows
\begin{align}\label{}
1-\frac{1}{\lambda_{\min,k}} \le eig\{ {{\bf I}_{{{\cal C}_k}}^{-1}}- ({\bf{\Sigma}}_{{{\cal C}_k}})^{-1} \}    \le 1-\frac{1}{\lambda_{\max,k}}
\end{align}
where $\lambda_{\min,k}$ and $\lambda_{\max,k}$  are the minimum and maximum eigenvalues of ${\bf{\Sigma}}_{{{\cal C}_k}}$ respectively. Then we obtain
\begin{align}\label{ICkeigen}
eig\big\{ {{\bf I}_{{{\cal C}_k}}^{-1}}- ({\bf{\Sigma}}_{{{\cal C}_k}})^{-1} \big\}\rightarrow 1 \quad \mbox{when} \quad {\lambda_{\min,k}}\rightarrow\infty.
\end{align}
By employing the Eigenvalue Interlacing Theorem \cite{horn_johnson_2012} which mainly states that the eigenvalues of the  principal submatrix can be bounded by eigenvalues of the whole matrix, we obtain for ${j \in {{\cal Q}_k}}$
\begin{align}\label{lambdapk}
\lambda_{\min,k}\le eig\big\{ {\bf{\Sigma}}_{{{\cal S}_j}} \big\} \le   \lambda_{\max,k}
\end{align}
where $\lambda_{\min,k}$ and $\lambda_{\max,k}$ are the minimum and maximum eigenvalues of the matrix ${\bf{\Sigma}}_{{{\cal C}_k}}$. Similarly, for $k=2,3,...,K$, we obtain
\begin{align}\label{lambdaqk}
\lambda_{\min,k}\le eig\big\{ {\bf{\Sigma}}_{{{\cal S}_k}} \big\} \le   \lambda_{\max,k}.
\end{align}
Then we can obtain
\begin{align}\label{ISjeigen}
eig\big\{ {{\bf I}_{{{\cal S}_j}}^{-1}}- ({\bf{\Sigma}}_{{{\cal S}_j}})^{-1} \big\}\rightarrow 1 \quad \mbox{when} \quad {\lambda_{\min,k}}\rightarrow\infty
\end{align}
for ${j \in {{\cal Q}_k}}$ and
\begin{align}\label{ISkeigen}
eig\big\{ {{\bf I}_{{{\cal S}_k}}^{-1}}- ({\bf{\Sigma}}_{{{\cal S}_k}})^{-1} \big\}\rightarrow 1 \quad \mbox{when} \quad {\lambda_{\min,k}}\rightarrow\infty
\end{align}
for $k=2,3,...,K$.

Weyl inequality (see Theorem 4.3.1 in \cite{horn_johnson_2012} for more details) states that let $A$ and $B$ be $n\times n$ Hermitian and let the respective eigenvalues of $A$, $B$, and $A+B$ be $\{\lambda_i(A)\}_{i=1}^n$, $\{\lambda_i(B)\}_{i=1}^n$, and $\{\lambda_i(A+B)\}_{i=1}^n$, each algebraically nondecreasing ordered, then
\begin{align}\label{Weylinequality1}
\lambda_i(A+B)\le \lambda_{i+j}(A) + \lambda_{n-j}(B),\qquad j=0,1,...,n-i
\end{align}
and
\begin{align}\label{Weylinequality2}
\lambda_{i-j+1}(A) + \lambda_{j}(B) \le \lambda_i(A+B),\qquad j=1,2,...,i
\end{align}
for each $i=1,2,...,n$. By employing (\ref{Weylinequality1}) and (\ref{Weylinequality2}), we can obtain that for $i=1,2,...,n$,
\begin{align}\label{boundsumeig}
\lambda_1(A) + \lambda_1(B) \le \lambda_i(A+B)\le  \lambda_n(A) + \lambda_n(B)
\end{align}
where $\lambda_1(A)$ and $\lambda_1(B)$ are the minimum eigenvalues of $A$ and $B$ respectively, and $\lambda_n(A)$ and $\lambda_n(B)$ are the maximum eigenvalues of $A$ and $B$ respectively. By employing the inequality in (\ref{boundsumeig}) into (\ref{Jk}), we obtain
\begin{align}
 &\bigg\{\min eig\big\{ {{\bf I}_{{{\cal C}_k}}^{-1}}- ({\bf{\Sigma}}_{{{\cal C}_k}})^{-1}\big\}\notag\\ \notag
 &\quad-\!\! \sum\limits_{j \in {{\cal Q}_k}}{\beta _j} \max eig\Big\{ \big[ {{\bf I}_{{{\cal S}_{j}}}^{-1}} - ({\bf{\Sigma}}_{{{\cal S}_{j}}})^{-1}\big]^{{{\cal C}_k}}\Big\}\notag\\
 &-\alpha_k\max eig\Big\{ \big[{{\bf I}_{{{\cal S}_{k}}}^{-1}} - ({\bf{\Sigma}}_{{{\cal S}_{k}}})^{-1}\big]^{{{\cal C}_k}}\Big\}\bigg\} \le eig\big\{ {\bf J}_k \big\}\notag\\
 &\le \bigg\{\max eig\big\{ {{\bf I}_{{{\cal C}_k}}^{-1}}- ({\bf{\Sigma}}_{{{\cal C}_k}})^{-1}\big\}\notag\\ \notag
 &\quad-\!\! \sum\limits_{j \in {{\cal Q}_k}}{\beta _j} \min eig\Big\{ \big[ {{\bf I}_{{{\cal S}_{j}}}^{-1}} - ({\bf{\Sigma}}_{{{\cal S}_{j}}})^{-1}\big]^{{{\cal C}_k}}\Big\}\notag\\
 &-\alpha_k\min eig\Big\{ \big[{{\bf I}_{{{\cal S}_{k}}}^{-1}} - ({\bf{\Sigma}}_{{{\cal S}_{k}}})^{-1}\big]^{{{\cal C}_k}}\Big\}\bigg\}.\label{JkEIGENbound}
\end{align}
Then by employing (\ref{ICkeigen}) and (\ref{ISjeigen})--(\ref{JkEIGENbound}), we obtain when ${\lambda_{\min,k}}\rightarrow\infty$, for $k=2,3,...,K$,
\begin{align} \label{Jkeigen}
\big(1- \sum\limits_{j \in {{\cal Q}_k}}{\beta _j} - \alpha_k\big) \le eig\big\{ {\bf J}_k \big\} \le 1.
\end{align}
This implies
\begin{align} \label{leftpositive}
{\bf J}_k - \big(1- \sum\limits_{j \in {{\cal Q}_k}}{\beta _j} - \alpha_k\big) {\bf I}_{{{\cal C}_k}} \succeq 0
\end{align}
and
\begin{align} \label{rightpositive}
{\bf I}_{{{\cal C}_k}} - {\bf J}_k   \succeq 0.
\end{align}

For $k=1$, by employing (\ref{ICkeigen}), (\ref{ISjeigen}) and ${\bf J}_1$ in (\ref{J1}), we obtain when ${\lambda_{\min,k}}\rightarrow\infty$,
\begin{align} \label{eigenJ1}
\big(1- \sum\limits_{j \in {{\cal Q}_1}}{\beta _j} \big) \le eig\big\{ {\bf J}_1 \big\} \le 1.
\end{align}

By employing the definition of ${\beta _k}$ in (\ref{beta_Theorem2}), it is clear that for all $k=2,3,...,K$,
\begin{align}\label{beta_larger_than_0}
{\beta _k} = {2^{K - k}}\gamma  > 0
\end{align}
where the positive-valued $\gamma$ is defined in (\ref{gamma_Theorem2}). Furthermore,
we can obtain
\begin{align}  \label{proof_Theorem_2_temp1}
\sum\limits_{j \in {{\cal Q}_1}} {{\beta _j}} &  \le \sum\limits_{j = 2}^K {{\beta _j}} \\  \notag
& = \sum\limits_{j = 2}^K {{2^{K - j}}\gamma } \\  \notag
& = \gamma \left( {{2^{K - 1}} - 1} \right)\\ \label{proof_Theorem_2_temp2}
& < 1,
\end{align}
where (\ref{proof_Theorem_2_temp1}) is a consequence of (\ref{Q_j_subset}) and (\ref{beta_larger_than_0}), and (\ref{proof_Theorem_2_temp2}) is from (\ref{gamma_Theorem2}). By employing (\ref{Q_j_subset}), (\ref{alpha_Theorem2}), (\ref{gamma_Theorem2}) and (\ref{beta_larger_than_0}), we obtain for all $j=2,3,...,K$,
\begin{align} \notag
{\alpha _k} + \sum\limits_{j \in {{\cal Q}_k}} {{\beta _j}}  & = 1 - {2^{K - k}}\gamma  + \sum\limits_{j \in {{\cal Q}_k}} {{\beta _j}} \\ \notag
& \le 1 - {2^{K - k}}\gamma  + \sum\limits_{j = k + 1}^K {{\beta _j}} \\ \notag
& = 1 - {2^{K - k}}\gamma  + \sum\limits_{j = k + 1}^K {{2^{K - j}}\gamma } \\ \notag
& = 1 - \gamma \\
& < 1.\label{sumlessthanone2}
\end{align}

Thus, by employing (\ref{Jkeigen})--(\ref{sumlessthanone2}), we can obtain for all $k=1,2,...,K$,
\begin{align}
{\bf J}_k \succ 0 \quad\mbox{when} \quad {\lambda_{\min,k}}\rightarrow\infty.
\end{align}

We can rewrite ${{L_k}({{\bf{x}}_{{{\cal C}_k}}})}$ under $H_1$ for $k=1,2,...,K$ in (\ref{sum_LLLR_local22}) as follows
\begin{align}
&{L_k}\left( {{{\bf{x}}_{{{\cal C}_k}}}} \right) ={{\bf {x}}^T_{{{\cal C}_K}}} {{ {{{\bf{J}}_k}} }} {{\bf {x}}_{{{\cal C}_K}}} -{e}_k\nonumber\\
&={\bf y}_{{{\cal C}_k}}^T {\bf{W}}_{{{\cal C}_k}}^T  {{ {{{\bf{J}}_k}} }} {\bf{W}}_{{{\cal C}_k}}{\bf y}_{{{\cal C}_k}}-{e}_k\label{changevariable}
\end{align}
where we use ${\bf{x}}_{{{\cal C}_k}} ={\bf{W}}_{{{\cal C}_k}} {\bf{y}}_{{{\cal C}_k}}$ to obtain (\ref{changevariable}) with ${\bf{y}}_{{{\cal C}_k}}\sim \mathcal{N}(0,{\bf{I}}_{{{\cal C}_k}})$ and a known whitening matrix ${\bf{W}}_{{{\cal C}_k}}$ satisfying the condition ${\bf{W}}_{{{\cal C}_k}}{\bf{W}}_{{{\cal C}_k}}^T={\bf{\Sigma}}_{{\cal C}_k}$. Let ${\bf{\Sigma}}_{{\cal C}_k}={\bf{V}}_{{{\cal C}_k}}^T diag\{\lambda_{0,k}, \lambda_{1,k},...,\lambda_{M_k-1,k}\}{\bf{V}}_{{{\cal C}_k}}$ be the eigenvalue/eigenvector decomposition for ${\bf{\Sigma}}_{{\cal C}_k}$, from (\ref{leftpositive}),
we can obtain for $k=2,3,...,K$,
\begin{align}
&{\bf y}_{{{\cal C}_k}}^T {\bf{W}}_{{{\cal C}_k}}^T  {{ {{{\bf{J}}_k}} }} {\bf{W}}_{{{\cal C}_k}}{\bf y}_{{{\cal C}_k}}\notag\\ \notag
&\ge \big(1- \sum\limits_{j \in {{\cal Q}_k}}{\beta _j} - \alpha_k\big) {\bf y}_{{{\cal C}_k}}^T {\bf{W}}_{{{\cal C}_k}}^T {\bf I}_{{{\cal C}_k}} {\bf{W}}_{{{\cal C}_k}} {\bf y}_{{{\cal C}_k}} \label{} \\
&=\big(1- \sum\limits_{j \in {{\cal Q}_k}}{\beta _j} - \alpha_k\big) {\bf y}_{{{\cal C}_k}}^T {\bf{\Sigma}}_{{\cal C}_k} {\bf y}_{{{\cal C}_k}} \label{signamMCda} \\
&= \big(1- \sum\limits_{j \in {{\cal Q}_k}}{\beta _j} - \alpha_k\big) {\bf y}_{{{\cal C}_k}}^T{\bf{V}}_{{{\cal C}_k}}^T diag\big\{\lambda_{0,k}, \lambda_{1,k},..., \notag\\ &\qquad\lambda_{M_k-1,k}\big\} {\bf{V}}_{{{\cal C}_k}}{\bf y}_{{{\cal C}_k}}\label{signamM2}\\
& = \big(1- \sum\limits_{j \in {{\cal Q}_k}}{\beta _j} - \alpha_k\big) \sum_{m=0}^{M_k-1} \lambda_{m,k} t_{m,{{\cal C}_k}}^2\label{signamM3}\\
&>\big(1- \sum\limits_{j \in {{\cal Q}_k}}{\beta _j} - \alpha_k\big) \lambda_{\max,k} t_{\max,{{\cal C}_k}}^2.\label{signamM5}
\end{align}
For ${\bf{\Sigma}}_{{\cal C}_k}$, the eigenvalues are defined as $\lambda_{0,k}, \lambda_{1,k},...,\lambda_{M_k-1,k}$. In going from (\ref{signamM2}) to (\ref{signamM3}), we define $t_{m,{{\cal C}_k}}$ for $m=0,1,...,M_k-1$ as the elements of the vector ${\bf{V}}_{{{\cal C}_k}}{\bf y}_{{{\cal C}_k}}$. In going from (\ref{signamM3}) to (\ref{signamM5}), we drop positive terms. Note that for $k=2,3,...,K$,
\begin{align}
&\lim_{\lambda_{\min,k}\rightarrow+\infty} \frac{1}{\lambda_{\max,k}-1}{\big(1- \sum\limits_{j \in {{\cal Q}_k}}{\beta _j} - \alpha_k\big) \lambda_{\max,k} t_{\max,{{\cal C}_k}}^2}\notag\\
&=\lim_{\lambda_{\min,k}\rightarrow+\infty} \big(1- \sum\limits_{j \in {{\cal Q}_k}}{\beta _j} - \alpha_k\big)t_{\max,{{\cal C}_k}}^2  >  0
\end{align}
which implies under $H_1$, for $k=2,3,...,K$,
\begin{align}\label{limitPdpositive}
\lim_{\lambda_{\min,k}\rightarrow+\infty}\frac{1}{\lambda_{\max,k}-1}{\bf y}_{{{\cal C}_k}}^T {\bf{W}}_{{{\cal C}_k}}^T  {{ {{{\bf{J}}_k}} }} {\bf{W}}_{{{\cal C}_k}}{\bf y}_{{{\cal C}_k}}  >  0.
\end{align}

By substituting (\ref{changevariable}) into (\ref{Pr_H1_LB_Lemma2}), we obtain for any given $\delta^{(1)}$, the detection probability $P_{D,k}(\delta^{(1)})$ in the $k$-th cluster for any $k=2,...,K$ is
\begin{align}
&\lim_{\lambda_{\min,k}\rightarrow+\infty}P_{D,k}(\delta^{(1)})\notag\\
&=\lim_{\lambda_{\min,k}\rightarrow+\infty}\Pr \Bigg( {\bf y}_{{{\cal C}_k}}^T {\bf{W}}_{{{\cal C}_k}}^T  {{ {{{\bf{J}}_k}} }} {\bf{W}}_{{{\cal C}_k}}{\bf y}_{{{\cal C}_k}}-{e}_k > \delta^{(1)} \bigg| {{H_1}} \Bigg)\label{PDfina1}\\
&=\lim_{\lambda_{\min,k}\rightarrow+\infty}\Pr \Bigg( \frac{1}{{\lambda_{\max,k}-1}}{\bf y}_{{{\cal C}_k}}^T {\bf{W}}_{{{\cal C}_k}}^T  {{ {{{\bf{J}}_k}} }} {\bf{W}}_{{{\cal C}_k}}{\bf y}_{{{\cal C}_k}}\notag\\
&\qquad\ > \frac{\delta^{(1)}}{{\lambda_{\max,k}-1}} + \frac{\ln\det({\bf{\Sigma}}_{{{\cal C}_k}})}{{\lambda_{\max,k}-1}} - \frac{\sum\limits_{j \in {{\cal Q}_k}}{\beta _j} \ln\det({\bf{\Sigma}}_{{{\cal S}_{j}}})}{{\lambda_{\max,k}-1}}\notag\\
&\qquad\ -\frac{\alpha_k\ln\det({\bf{\Sigma}}_{{{\cal S}_k}})}{{\lambda_{\max,k}-1}} \bigg| {{H_1}}  \Bigg) \label{PDfina2}\\
&=\lim_{\lambda_{\min,k}\rightarrow+\infty}\Pr \Bigg( \frac{1}{{\lambda_{\max,k}-1}}{\bf y}_{{{\cal C}_k}}^T {\bf{W}}_{{{\cal C}_k}}^T  {{ {{{\bf{J}}_k}} }} {\bf{W}}_{{{\cal C}_k}}{\bf y}_{{{\cal C}_k}}\notag\\
&\qquad\ > 0\ \bigg| {{H_1}}  \Bigg)\label{PDfina3}\\
&=1.\label{PDfina4}
\end{align}
In going from (\ref{PDfina1}) to (\ref{PDfina2}), we employ $e_k$ in (\ref{Mk}) and divide $(\lambda_{\max,k}-1)$ on both sides for large $\lambda_{\max,k}>1$. In going from (\ref{PDfina2}) to (\ref{PDfina3}), $\delta^{(1)}/(\lambda_{\max,k}-1)\rightarrow 0$, and $\ln\det({\bf{\Sigma}}_{{{\cal C}_k}})/(\lambda_{\max,k}-1)\rightarrow 0$ since $\ln\det({\bf{\Sigma}}_{{{\cal C}_k}})$ equals to the logarithm of the product of $eig\{{\bf{\Sigma}}_{{{\cal C}_k}}\}$ which is also upper bounded by $\ln M_k\lambda_{\max,k}$. Similarly, by employing (\ref{lambdapk}) and (\ref{lambdaqk}), we can obtain $\ln\det({\bf{\Sigma}}_{{{\cal S}_k}})/(\lambda_{\max,k}-1)\rightarrow 0$ and $\ln\det({\bf{\Sigma}}_{{{\cal S}_j}})/(\lambda_{\max,k}-1)\rightarrow 0$ which lead to (\ref{PDfina3}). The result in (\ref{PDfina4}) follows by employing (\ref{limitPdpositive}).

For $k=1$, by employing similar steps as used in (\ref{signamMCda})--(\ref{signamM3}), we obtain
\begin{align}
&{\bf y}_{{{\cal C}_1}}^T {\bf{W}}_{{{\cal C}_1}}^T  {{ {{{\bf{J}}_1}} }} {\bf{W}}_{{{\cal C}_1}}{\bf y}_{{{\cal C}_1}}\notag\\
&>\big(1- \sum\limits_{j \in {{\cal Q}_1}}{\beta _j} \big) \lambda_{\max,1} t_{\max,{{\cal C}_1}}^2.\label{signamM4}
\end{align}
Then similar to (\ref{limitPdpositive}), we can obtain
\begin{align}\label{Pd1positive}
\lim_{\lambda_{\min,1}\rightarrow+\infty}\frac{1}{\lambda_{\max,1}-1}{\bf y}_{{{\cal C}_1}}^T {\bf{W}}_{{{\cal C}_1}}^T  {{ {{{\bf{J}}_1}} }} {\bf{W}}_{{{\cal C}_1}}{\bf y}_{{{\cal C}_1}}  >  0.
\end{align}
Thus, we obtain for any given $\delta^{(1)}$, the detection probability $P_{D,1}(\delta^{(1)})$ in the first   cluster is
\begin{align}
&\lim_{\lambda_{\min,1}\rightarrow+\infty}P_{D,1}(\delta^{(1)})\notag\\
&=\lim_{\lambda_{\min,1}\rightarrow+\infty}\Pr \Bigg( \frac{1}{{\lambda_{\max,1}-1}}{\bf y}_{{{\cal C}_1}}^T {\bf{W}}_{{{\cal C}_1}}^T  {{ {{{\bf{J}}_1}} }} {\bf{W}}_{{{\cal C}_1}}{\bf y}_{{{\cal C}_1}}\notag\\
&\ > \frac{\delta^{(1)}}{{\lambda_{\max,1}-1}} + \frac{\ln\det({\bf{\Sigma}}_{{{\cal C}_1}})}{{\lambda_{\max,1}-1}} - \frac{\sum\limits_{j \in {{\cal Q}_1}}{\beta _j} \ln\det({\bf{\Sigma}}_{{{\cal S}_{j}}})}{{\lambda_{\max,1}-1}} \bigg| {{H_1}}  \Bigg) \label{PD1fina2}\\
&=\lim_{\lambda_{\min,1}\rightarrow+\infty}\Pr \Bigg( \frac{1}{{\lambda_{\max,1}-1}}{\bf y}_{{{\cal C}_1}}^T {\bf{W}}_{{{\cal C}_1}}^T  {{ {{{\bf{J}}_1}} }} {\bf{W}}_{{{\cal C}_1}}{\bf y}_{{{\cal C}_1}}\notag\\
&\qquad\ > 0\ \bigg| {{H_1}}  \Bigg)\label{PD1fina3}\\
&=1.\label{PD1fina4}
\end{align}
where we employ ${e}_1$ in (\ref{e1}) to obtain (\ref{PD1fina2}). (\ref{PD1fina4}) is a consequence of (\ref{Pd1positive}).
Then by employing (\ref{PDfina4}) and (\ref{PD1fina4}) together, finally we can show that for all $k=1,2,...,K$
\begin{align}\label{PDallk}
\lim_{\lambda_{\min,k}\rightarrow+\infty}P_{D,k}(\delta^{(1)}) = 1.
\end{align}
Note that ${\lambda_{\min,k}\rightarrow+\infty}$ for all $k=1,2,...,K$ if and only if the minimum among them $\lambda_{\min}\rightarrow+\infty$. Finally, we obtain for all $k=1,2,...,K$
\begin{align}\label{PDminall}
\lim_{\lambda_{\min}\rightarrow+\infty}P_{D,k}(\delta^{(1)}) = 1.
\end{align}

Now we consider the case under $H_0$ where ${\bf{x}}_{{{\cal C}_k}}\sim \mathcal{N}(0,{\bf{I}}_{{{\cal C}_k}})$ for $k=1,2,...,K$. From (\ref{rightpositive}), we obtain
\begin{align}\label{sumyCk}
{{\bf {x}}^T_{{{\cal C}_K}}}   {{ {{{\bf{J}}_k}} }} {{\bf {x}}_{{{\cal C}_K}}}  \le {{\bf {x}}^T_{{{\cal C}_K}}} {\bf I}_{{{{\cal C}_K}}} {{\bf {x}}_{{{\cal C}_K}}}
=\sum_{m=0}^{M_k-1} {x}_{{m,{\cal C}_k}}^2
\end{align}
where we  employ the second inequalities in (\ref{Jkeigen}) and (\ref{eigenJ1}) to obtain the first inequality in (\ref{sumyCk}). The equality in (\ref{sumyCk}) follows since ${x}_{{m,{\cal C}_k}}$ is the m-th element of the vector ${\bf x}_{{{\cal C}_k}}\sim \mathcal{N}(0,{\bf{I}}_{{{\cal C}_k}})$ under $H_0$.

Before we show $\lim_{\lambda_{\min,k}\rightarrow+\infty}P_{f,k}(\delta^{(0)})=0$ for $k=2,3,...,K$, we first let
\begin{align}\label{SigamSj}
{\bf{\Sigma}}_{{{\cal C}_k}} \buildrel \Delta \over  = \left[ {\begin{array}{*{20}{c}}
	{{\bf{\Xi }}}_{k}&{{\bf{\Omega }}}_{k}\\
	{{\bf{\Omega }}}_k^T& {{\bf{\Sigma}}_{{{\cal S}_{j}}}}
	\end{array}} \right]
\end{align}
where ${{\bf{\Sigma}}_{{{\cal S}_{j}}}}$ is the covariance matrix of the separator set for $j\in{{\cal Q}_k}$. By employing the Schur complement \cite{horn_johnson_2012}, we obtain
\begin{small}
\begin{align}\label{Schurcompl}
&\big({\bf{\Sigma}}_{{{\cal C}_k}}\big)^{-1} = \notag\\
&\!\!\!\left[\!\!\!\! {\begin{array}{*{20}{c}}
	\!\!\!\!\big( {{\bf{\Xi }}}_{k} - {{\bf{\Omega }}}_{k}{{\bf{\Sigma}}_{{{\cal S}_{j}}}^{-1}}{{\bf{\Omega }}}_k^T \big)^{-1}\!\!&\!\!\!\!-{{\bf{\Xi }}}_{k}^{-1}{{\bf{\Omega }}}_k\big({{\bf{\Sigma}}_{{{\cal S}_{j}}}} - {{\bf{\Omega }}}_k^T {{\bf{\Xi }}}_{k}^{-1} {{\bf{\Omega }}}_k \big)^{-1}\\
	-\big({{\bf{\Sigma}}_{{{\cal S}_{j}}}} - {{\bf{\Omega }}}_k^T {{\bf{\Xi }}}_{k}^{-1} {{\bf{\Omega }}}_k \big)^{-1}{{\bf{\Omega }}}_k^T{{\bf{\Xi }}}_{k}^{-1} \!\!&\!\! \big({{\bf{\Sigma}}_{{{\cal S}_{j}}}} - {{\bf{\Omega }}}_k^T {{\bf{\Xi }}}_{k}^{-1} {{\bf{\Omega }}}_k \big)^{-1}
	\end{array}} \!\!\!\!\right].
\end{align}
\end{small}Since it is clear that $eig\{({\bf{\Sigma}}_{{{\cal C}_k}})^{-1}\}\rightarrow 0$ when $\lambda_{\min,k}\rightarrow+\infty$, then by employing the Eigenvalue Interlacing Theorem\footnote{The Eigenvalue Interlacing Theorem states that the eigenvalues of the principal submatrix $\bf B$ obtained by deleting both $i$-th row and $i$-th column of the matrix $\bf A$ for some values of $i$ can be bounded by the eigenvalues of the matrix $\bf A$ \cite{horn_johnson_2012}.}, we obtain the eigenvalues of the principal submatrix $eig\{\big( {{\bf{\Xi }}}_{k} - {{\bf{\Omega }}}_{k} ({{\bf{\Sigma}}_{{{\cal S}_{j}}}})^{-1} {{\bf{\Omega }}}_k^T \big)^{-1}\}$ (which are bounded by $eig\{({\bf{\Sigma}}_{{{\cal C}_k}})^{-1}\}$) must approach zero. Thus $eig\{( {{\bf{\Xi }}}_{k} - {{\bf{\Omega }}}_{k} ({{\bf{\Sigma}}_{{{\cal S}_{j}}}})^{-1} {{\bf{\Omega }}}_k^T )\}\rightarrow +\infty$. By employing a property of the Schur complement \cite{horn_johnson_2012}, we obtain when $\lambda_{\min,k}\rightarrow+\infty$
\begin{align}\label{SchurCompp}
\frac{det({\bf{\Sigma}}_{{{\cal C}_k}})}{det({{\bf{\Sigma}}_{{{\cal S}_{j}}}})} = det({{\bf{\Xi }}}_{k} - {{\bf{\Omega }}}_{k} ({{\bf{\Sigma}}_{{{\cal S}_{j}}}})^{-1} {{\bf{\Omega }}}_k^T )\rightarrow+\infty.
\end{align}
(\ref{SchurCompp}) implies that when $\lambda_{\min,k}\rightarrow+\infty$,
 \begin{align}\label{SjComplem}
 \big(\ln\det({\bf{\Sigma}}_{{{\cal C}_k}})-\ln\det({\bf{\Sigma}}_{{{\cal S}_{j}}})\big)\rightarrow+\infty.
 \end{align}
Similar to the steps taken from (\ref{SigamSj}) to (\ref{SchurCompp}), we can also obtain when $\lambda_{\min,k}\rightarrow+\infty$,
 \begin{align}\label{SkComplem}
\big(\ln\det({\bf{\Sigma}}_{{{\cal C}_k}})-\ln\det({\bf{\Sigma}}_{{{\cal S}_{k}}})\big)\rightarrow+\infty.
 \end{align}

Next we will show $\lim_{\lambda_{\min,k}\rightarrow+\infty}P_{f,k}(\delta^{(0)})=0$ for $k=2,3,...,K$. By substituting ${{L_k}({{\bf{x}}_{{{\cal C}_k}}})}$ in (\ref{sum_LLLR_local22}) into (\ref{Pr_H0_LB_Lemma2}), we obtain for any given $\delta^{(0)}$, the false alarm probability $P_{f,k}(\delta^{(0)})$ in the $k$-th cluster for any $k=2,...,K$ is
\begin{align}
&\lim_{\lambda_{\min,k}\rightarrow+\infty}P_{f,k}(\delta^{(0)}) \notag\\
&= \lim_{\lambda_{\min,k}\rightarrow+\infty}\Pr \Bigg( {{\bf {x}}^T_{{{\cal C}_K}}}   {{ {{{\bf{J}}_k}} }} {{\bf {x}}_{{{\cal C}_K}}}-{e}_k>   \delta^{(0)}
\bigg| {{H_0}}\!\! \Bigg)\label{pflimitk1}\\
&\le  \lim_{\lambda_{\min,k}\rightarrow+\infty}\Pr \Bigg( \sum_{m=0}^{M_k-1} {x}_{{m,{\cal C}_k}}^2 >   \delta^{(0)} + \ln\det({\bf{\Sigma}}_{{{\cal C}_k}}) \notag \\
&\qquad -\sum\limits_{j \in {{\cal Q}_k}}{\beta _j} \ln\det({\bf{\Sigma}}_{{{\cal S}_{j}}})-\alpha_k\ln\det({\bf{\Sigma}}_{{{\cal S}_{k}}})\bigg| {{H_0}}\!\! \Bigg)\label{pflimitk2}\\
&\le  \lim_{\lambda_{\min,k}\rightarrow+\infty}\Pr \Bigg(\!\! \sum_{m=0}^{M_k-1} {x}_{{m,{\cal C}_k}}^2 >   \delta^{(0)}\! + \!\!{\sum\limits_{j \in {{\cal Q}_k}}{\beta _j}}\cdot\Big(\!\ln\det({\bf{\Sigma}}_{{{\cal C}_k}}) \notag \\
&\qquad - \ln\det({\bf{\Sigma}}_{{{\cal S}_{j}}})\Big) + \alpha_k \cdot \Big(  \ln\det({\bf{\Sigma}}_{{{\cal C}_k}})\notag \\
&\qquad-\ln\det({\bf{\Sigma}}_{{{\cal S}_{k}}})\Big) \bigg| {{H_0}}\!\! \Bigg)\label{pflimitk3}\\
&=\lim_{\lambda_{\min,k}\rightarrow+\infty}\Pr \Bigg( \sum_{m=0}^{M_k-1} {x}_{{m,{\cal C}_k}}^2 >  +\infty\bigg| {{H_0}}\Bigg)\label{pflimitk4}\\
&=0.\label{pflimitk5}
\end{align}
In going from (\ref{pflimitk1}) to (\ref{pflimitk2}), we employ $e_k$ in (\ref{Mk}) and the inequality in (\ref{sumyCk}). In going from (\ref{pflimitk2}) to (\ref{pflimitk3}), we employ (\ref{sumlessthanone2}) to lower bound $\ln\det({\bf{\Sigma}}_{{{\cal C}_k}})$ by the sum of $\sum_{j \in {{\cal Q}_k}}{\beta _j}\ln\det({\bf{\Sigma}}_{{{\cal C}_k}})$ and $\alpha_{k}\ln\det({\bf{\Sigma}}_{{{\cal C}_k}})$. By employing the results in (\ref{SjComplem}) and (\ref{SkComplem}), $\delta^{(0)}\! + \!\!{\sum_{j \in {{\cal Q}_k}}{\beta _j}}\cdot\big(\!\ln\det({\bf{\Sigma}}_{{{\cal C}_k}}) - \ln\det({\bf{\Sigma}}_{{{\cal S}_{j}}})\big) + \alpha_k \cdot \big(  \ln\det({\bf{\Sigma}}_{{{\cal C}_k}})-\ln\det({\bf{\Sigma}}_{{{\cal S}_{k}}})\big)$ in (\ref{pflimitk3}) approaches positive infinity under the limit. (\ref{pflimitk5}) follows since $\sum_{m=0}^{M_k-1} {x}_{{m,{\cal C}_k}}^2$ can not be larger than the positive infinity.

By substituting ${{L_1}\left( {{{\bf{x}}_{{{\cal C}_1}}}} \right) }$ in (\ref{sum_LLLR_local22}) into (\ref{Pr_H0_LB_Lemma2}) and then taking similar steps as those taken to go from (\ref{pflimitk1}) to (\ref{SigamSj}), we can also show that
\begin{align}\label{firstclusterH0}
\lim_{\lambda_{\min,1}\rightarrow+\infty}P_{f,1}(\delta^{(0)}) \le 0,
\end{align}
for the first cluster. Then by using the fact that the false alarm probability $P_{f,k}(\delta^{(0)})$ is non-negative and employing (\ref{pflimitk5}) and (\ref{firstclusterH0}) together, finally we can show that for all $k=1,2,...,K$,
\begin{align}\label{firstresultPfH0}
\lim_{\lambda_{\min,k}\rightarrow+\infty}P_{f,k}(\delta^{(0)}) = 0.
\end{align}
Again, ${\lambda_{\min,k}\rightarrow+\infty}$ for all $k=1,2,...,K$ if and only if $\lambda_{\min}\rightarrow+\infty$. Then we obtain for all $k=1,2,...,K$
\begin{align}\label{PfH0allK}
\lim_{\lambda_{\min}\rightarrow+\infty}P_{f,k}(\delta^{(0)}) = 0.
\end{align}

Using the results in (\ref{PDminall}) and (\ref{PfH0allK}), the limiting behavior of the lower bound on the average number of transmissions saved by ordering the global communications in (\ref{Theorem_state_K_s}) for $K>1$ becomes
\begin{align}\label{TheoremstateKslimit}
 {K_s}  \notag
& > \left( {\left\lceil {\frac{K}{2}} \right\rceil  - 1} \right){\pi _1}\cdot K
 + \left( {\left\lceil {\frac{K}{2}} \right\rceil  - 1} \right){\pi _0}\cdot K\nonumber\\
&\qquad\ - \left( {\left\lceil {\frac{K}{2}} \right\rceil  - 1} \right)\left( {K - 1} \right)\nonumber\\
&  ={\left\lceil {\frac{K}{2}} \right\rceil  - 1}.
\end{align}

\bibliographystyle{IEEEtran}
\bibliography{refs}

\begin{thebibliography}{10}
\providecommand{\url}[1]{#1}
\csname url@samestyle\endcsname
\providecommand{\newblock}{\relax}
\providecommand{\bibinfo}[2]{#2}
\providecommand{\BIBentrySTDinterwordspacing}{\spaceskip=0pt\relax}
\providecommand{\BIBentryALTinterwordstretchfactor}{4}
\providecommand{\BIBentryALTinterwordspacing}{\spaceskip=\fontdimen2\font plus
\BIBentryALTinterwordstretchfactor\fontdimen3\font minus
  \fontdimen4\font\relax}
\providecommand{\BIBforeignlanguage}[2]{{%
\expandafter\ifx\csname l@#1\endcsname\relax
\typeout{** WARNING: IEEEtran.bst: No hyphenation pattern has been}%
\typeout{** loaded for the language `#1'. Using the pattern for}%
\typeout{** the default language instead.}%
\else
\language=\csname l@#1\endcsname
\fi
#2}}
\providecommand{\BIBdecl}{\relax}
\BIBdecl

\bibitem{akyildiz2002wireless}
I.~F. Akyildiz, W.~Su, Y.~Sankarasubramaniam, and E.~Cayirci, ``Wireless sensor
  networks: a survey,'' \emph{Computer networks}, vol.~38, no.~4, pp. 393--422,
  2002.

\bibitem{appadwedula2005energy}
S.~Appadwedula, V.~V. Veeravalli, and D.~L. Jones, ``Energy-efficient detection
  in sensor networks,'' \emph{IEEE Journal on Selected areas in
  communications}, vol.~23, no.~4, pp. 693--702, 2005.

\bibitem{6047550}
M.~Noori and M.~Ardakani, ``Energy efficiency of universal decentralized
  estimation in random sensor networks,'' \emph{IEEE Transactions on Wireless
  Communications}, vol.~10, no.~12, pp. 4023--4028, December 2011.

\bibitem{4545248}
R.~S. Blum and B.~M. Sadler, ``Energy efficient signal detection in sensor
  networks using ordered transmissions,'' \emph{IEEE Transactions on Signal
  Processing}, vol.~56, no.~7, pp. 3229--3235, July 2008.

\bibitem{zhang2017ordering}
J.~Zhang, Z.~Chen, R.~S. Blum, X.~Lu, and W.~Xu, ``Ordering for reduced
  transmission energy detection in sensor networks testing a shift in the mean
  of a gaussian graphical model,'' \emph{IEEE Transactions on Signal
  Processing}, vol.~65, no.~8, pp. 2178--2189, 2017.

\bibitem{wei2012distributed}
C.~Wei, A.~Wiesel, and R.~S. Blum, ``Distributed change detection in gaussian
  graphical models,'' in \emph{Information Sciences and Systems (CISS), 2012
  46th Annual Conference on}.\hskip 1em plus 0.5em minus 0.4em\relax IEEE,
  2012, pp. 1--4.

\bibitem{aittomaki2011resource}
T.~Aittom{\"a}ki, H.~Godrich, H.~V. Poor, and V.~Koivunen, ``Resource
  allocation for target detection in distributed mimo radars,'' in
  \emph{Signals, Systems and Computers (ASILOMAR), 2011 Conference Record of
  the Forty Fifth Asilomar Conference on}.\hskip 1em plus 0.5em minus
  0.4em\relax IEEE, 2011, pp. 873--877.

\bibitem{bilmes2005graphical}
J.~A. Bilmes and C.~Bartels, ``Graphical model architectures for speech
  recognition,'' \emph{IEEE signal processing magazine}, vol.~22, no.~5, pp.
  89--100, 2005.

\bibitem{wiesel2009decomposable}
A.~Wiesel and A.~O. Hero, ``Decomposable principal component analysis,''
  \emph{IEEE Transactions on Signal Processing}, vol.~57, no.~11, pp.
  4369--4377, 2009.

\bibitem{wille2004sparse}
A.~Wille, P.~Zimmermann, E.~Vranov{\'a}, A.~F{\"u}rholz, O.~Laule, S.~Bleuler,
  L.~Hennig, A.~Preli{\'c}, P.~von Rohr, L.~Thiele \emph{et~al.}, ``Sparse
  graphical gaussian modeling of the isoprenoid gene network in arabidopsis
  thaliana,'' \emph{Genome biology}, vol.~5, no.~11, p. R92, 2004.

\bibitem{willsky2002multiresolution}
A.~S. Willsky, ``Multiresolution markov models for signal and image
  processing,'' \emph{Proceedings of the IEEE}, vol.~90, no.~8, pp. 1396--1458,
  2002.

\bibitem{meng2013distributed}
Z.~Meng, D.~Wei, A.~Wiesel, and A.~Hero~III, ``Distributed learning of gaussian
  graphical models via marginal likelihoods,'' in \emph{Artificial Intelligence
  and Statistics}, 2013, pp. 39--47.

\bibitem{mohan2014node}
K.~Mohan, P.~London, M.~Fazel, D.~Witten, and S.-I. Lee, ``Node-based learning
  of multiple gaussian graphical models,'' \emph{The Journal of Machine
  Learning Research}, vol.~15, no.~1, pp. 445--488, 2014.

\bibitem{cetin2006distributed}
M.~Cetin, L.~Chen, J.~W. Fisher, A.~T. Ihler, R.~L. Moses, M.~J. Wainwright,
  and A.~S. Willsky, ``Distributed fusion in sensor networks,'' \emph{IEEE
  Signal Processing Magazine}, vol.~23, no.~4, pp. 42--55, 2006.

\bibitem{guestrin2004distributed}
C.~Guestrin, P.~Bodik, R.~Thibaux, M.~Paskin, and S.~Madden, ``Distributed
  regression: an efficient framework for modeling sensor network data,'' in
  \emph{Proceedings of the 3rd international symposium on Information
  processing in sensor networks}.\hskip 1em plus 0.5em minus 0.4em\relax ACM,
  2004, pp. 1--10.

\bibitem{he2011dependency}
M.~He and J.~Zhang, ``A dependency graph approach for fault detection and
  localization towards secure smart grid,'' \emph{IEEE Transactions on Smart
  Grid}, vol.~2, no.~2, pp. 342--351, 2011.

\bibitem{electric6687941}
Y.~Weng, R.~Negi, and M.~D. Ilic, ``Graphical model for state estimation in
  electric power systems,'' in \emph{2013 IEEE International Conference on
  Smart Grid Communications (SmartGridComm)}, Oct 2013, pp. 103--108.

\bibitem{abbasi2007survey}
A.~A. Abbasi and M.~Younis, ``A survey on clustering algorithms for wireless
  sensor networks,'' \emph{Computer communications}, vol.~30, no. 14-15, pp.
  2826--2841, 2007.

\bibitem{lauritzen1996graphical}
S.~L. Lauritzen, \emph{Graphical models}.\hskip 1em plus 0.5em minus
  0.4em\relax Clarendon Press, 1996, vol.~17.

\bibitem{horn_johnson_2012}
R.~A. Horn and C.~R. Johnson, \emph{Matrix Analysis}, 2nd~ed.\hskip 1em plus
  0.5em minus 0.4em\relax Cambridge University Press, 2012.

\end{thebibliography}

\end{document}